\documentclass[a4paper,11pt]{article}
\usepackage{jcappub}
\usepackage{amsmath}
\usepackage{graphicx}
\usepackage{latexsym}
\usepackage{xspace}
\usepackage{color}
\usepackage{bm}
\usepackage{relsize}
\usepackage{tabularx}
\usepackage{multirow}
\usepackage{amssymb}
\usepackage[table]{xcolor}

\usepackage{braket}
\usepackage{comment}
\usepackage[utf8]{inputenc}
\usepackage{slashed}
\usepackage{empheq}

\usepackage{subfigure}
\usepackage{here}
\usepackage[colorlinks=true, urlcolor=blue, citecolor=blue]{hyperref}
\usepackage[ruled,vlined]{algorithm2e}

\makeatletter
\gdef\@fpheader{}
\g@addto@macro\bfseries{\boldmath}
\makeatother

\newcommand{\prn}[1]{\left( {#1} \right)}
\newcommand{\com}[1]{\left[ {#1} \right]}

\newcommand{\abs}[1]{\left\vert {#1} \right\vert}

\newcommand{\dif}[2]{\frac{\mathrm{d} #1}{\mathrm{d} #2}}
\newcommand{\pdif}[2]{\frac{\partial #1}{\partial #2}}

\newcommand{\uir}{\mathrm{IR}}
\newcommand{\uuv}{\mathrm{UV}}

\newcommand{\ie}{\textsl{i.e.~}}

\newcommand{\dd}{\mathrm{d}}

\newcommand{\sss}[1]{{\scriptscriptstyle{#1}}}

\newcommand{\uPl}{\mathrm{Pl}}

\newcommand{\usssPl}{\sss{\uPl}}

\newcommand{\mpl}{M_\usssPl}

\newcommand{\efolds}{$e$-folds}
\newcommand{\efold}{$e$-fold}

\newcommand{\beq}{\begin{equation}}
\newcommand{\eeq}{\end{equation}}
\newcommand{\bea}{\begin{equation}\begin{aligned}}
\newcommand{\eea}{\end{aligned}\end{equation}}

\newlength{\wsingfig}
\newlength{\wdblefig}
\newlength{\wquadfig}
\newlength{\wtriplefig}
\newcommand{\Eq}[1]{Eq.~(\ref{#1})}
\newcommand{\Eqs}[1]{Eqs.~(\ref{#1})}

\newcommand{\Sec}[1]{Sec.~\ref{#1}}

\subheader{}

\title{
A consistent formulation of stochastic inflation II: Reassessing its role in nonlinear cosmological perturbations
}

\author[a]{Diego Cruces,}
\author[b,c]{Tomotaka Kuroda}

\affiliation[a]{Institute of Theoretical Physics, Chinese Academy of Sciences, Beijing 100190, China}
\affiliation[b]{Department of Physics, Institute of Science Tokyo, Tokyo, 152-8551, Japan}
\affiliation[c]{Cosmology, Gravity and Astroparticle Physics Group, Center for Theoretical Physics of the Universe, Institute for Basic Science (IBS), Daejeon, 34126, Korea}

\emailAdd{dcruces@itp.ac.cn}
\emailAdd{kuroda.t.ad@m.titech.ac.jp}

\date{today}

\begin{document}
\sloppy

\abstract{We question whether the conventional formulation of stochastic inflation can provide a quantitatively reliable description of cosmological perturbations. {More fundamentally,} we reassess the role and motivations of the stochastic approach to study cosmological perturbations, distinct from its well-founded motivation for resumming IR secular divergences. To provide a quantitative basis for this reassessment, we extend stochastic inflation beyond the linear regime by incorporating non-Markovian effects arising from infrared–ultraviolet interactions and terms quadratic in ultraviolet fluctuations, which generate non-white noise. We derive three-point correlation functions for a spectator scalar field in de Sitter space and a slow-roll inflaton including metric perturbations, demonstrating exact agreement with in-in quantum-field-theoretic predictions for superhorizon nonlinear evolution. Our results show that both non-Markovian effects and quadratic (higher order) noise terms are essential for quantitative accuracy, implying that the conventional stochastic-inflation framework is, at best, quantitatively justified only at linear order.}

\arxivnumber{}

\maketitle


\flushbottom


\section{Introduction}\label{Sec:1}

Cosmological inflation, a period of accelerated expansion in the very early universe, could give rise to quantum fluctuations that explain several observed phenomena, such as the temperature anisotropies of the cosmic microwave background (CMB). Thus, establishing theories that describe the quantum dynamics of the inflationary background, or (quasi-)de Sitter spacetime, is crucial for deriving reliable predictions for cosmological observables from first principles.

In this paper, we employ the stochastic formalism~\cite{Starobinsky:1986fx,Starobinsky:1994bd} as a theoretical framework for describing quantum perturbations in this setting. The stochastic formalism can be regarded as an effective field theory for infrared (IR) degrees of freedom, in which the effects of ultraviolet (UV) fluctuations are systematically incorporated through stochastic noise~\cite{PerreaultLevasseur:2013kfq,Burgess:2014eoa,Burgess:2015ajz,Moss:2016uix,Grain:2017dqa,Gorbenko:2019rza,Baumgart:2019clc,Mirbabayi:2019qtx,Mirbabayi:2020vyt,Cohen:2020php,Cohen:2021fzf,Cruces:2018cvq,Cruces:2021iwq,Cruces:2022imf,Launay:2024qsm,Li:2025azq,Ye:2026saa,Li:2026lwl}. At a heuristic level, this formalism offers a rough intuitive picture of field dynamics in terms of classical stochastic evolution rather than a direct quantum-field-theoretic calculation: the fields evolve under the influence of the potential and Hubble friction while receiving stochastic kicks originating from quantum fluctuations \cite{Starobinsky:1986fx,Nakao:1988yi,Sasaki:1987gy,Sasaki:1988df,Nambu:1987ef,Nambu:1988je}. 

The stochastic formalism is particularly well suited to treating the infrared divergences that arise for light spectator fields in de Sitter spacetime, which pose a well-known challenge to conventional perturbative methods. Within the stochastic approach, the relevant IR loop contributions are effectively resummed, and the secular growth appearing in perturbation theory is reinterpreted as the growth of statistical variances~\cite{Tsamis:2005hd,Enqvist:2008kt,Seery:2010kh,Garbrecht:2013coa,Garbrecht:2014dca,Tokuda:2017fdh,Tokuda:2018eqs,Honda:2023unh,Cespedes:2023aal,Palma:2023idj,Palma:2023uwo,Palma:2025oux}. Moreover, the formalism allows one to compute probability distribution functions (PDFs) directly. Since PDFs encode information contained in correlation functions of arbitrarily high order~\cite{Nakao:1988yi,Finelli:2008zg,Markkanen:2019kpv,Markkanen:2020bfc,Cable:2020dke,Cable:2022uwd,Cable:2023gdz,Bhattacharya:2022wjl,Honda:2024evc, Tokeshi:2024kuv, Takahashi:2025hqt, Nasuda:2025gyd, Takahashi:2026foq}, this provides a substantial technical advantage over perturbative approaches, in which calculations typically require truncation at finite order.

Moreover, the stochastic formalism has recently been considered applicable to the study of cosmological perturbations in nonlinear and even genuinely nonperturbative regimes. One representative example is the stochastic-$\delta N$ formalism~\cite{Fujita:2013cna,Fujita:2014tja,Vennin:2015hra} and its applications to various models~\cite{Vennin:2015hra,Pattison:2017mbe,Ezquiaga:2019ftu,Ballesteros:2020sre,Figueroa:2020jkf,Figueroa:2021zah,Tomberg:2023kli,Mishra:2023lhe,Tada:2023fvd,Mizuguchi:2024kbl,Murata:2025onc,Kuroda:2025coa,Raatikainen:2025gpd,Ando:2020fjm,Tada:2021zzj,Kitajima:2021fpq,Hooshangi:2021ubn,Gow:2022jfb,Raatikainen:2023bzk,Animali:2024jiz,Vennin:2024yzl,Jackson:2024aoo,Animali:2025pyf,Choudhury:2025kxg,Murata:2026yqb,Saha:2026cay,Animali:2026rvi,Animali:2026omc,Tomberg:2026uxl,Blachier:2025tcq,Blachier:2025iwk}. Such studies are motivated by phenomena such as primordial black holes (PBHs) and scalar-induced gravitational waves. In this context, it is therefore crucial to ensure that the formulation can properly describe the nonlinear regime in the amplitude of perturbations over inflationary timescales. However, we show in this paper that the conventional formulation cannot, as it stands, provide a quantitatively accurate description of nonlinear cosmological perturbations. Consequently, its characterization of non-Gaussianity, including the tail behavior of the probability distribution, remains quantitatively incomplete, regardless of any qualitative validity it might retain.

In particular, stochastic-$\delta N$-type approaches employ the Langevin equations governing inflaton dynamics and the corresponding Fokker--Planck equations. We identify the following two key aspects that require reconsideration for a quantitatively accurate description of nonlinear perturbations. {In what follows, we refer to formulations that neglect either or both of these aspects as the ``conventional formulation''.}\footnote{{For instance, as we demonstrate in the following sections, if the noise is assumed to take the conventional form $\Xi=H(\phi_{\mathrm{IR}})/(2\pi)\xi$, with normalized Gaussian noise $\xi$, as is commonly done in the literature, non-Markovian effects and quadratic (and higher-order) noise contributions are neglected, leading to quantitatively inaccurate predictions for nonlinear perturbations.}}
\begin{description}
	\item[Non-Markovian effects]\mbox{} \\
    First, some works still assume Markovian dynamics. For instance, the stochastic-$\delta N$ formalism employs the ``first-passage-time formulation" to compute the \efold. However, this assumption is not generally guaranteed to be valid for describing nonlinear cosmological perturbations. This is because the evolution of UV modes on top of the IR stochastic dynamics can render the resulting system nonlocal in time, leading to non-Markovian effects. We can observe such non-Markovianity from the simple discussion to considering the time scale of the system (IR modes) and the environment (UV modes) from the open quantum system perspective. For instance, UV modes oscillate rapidly deep inside the horizon, while the IR modes can be treated as approximately constant. Thus, there is a huge time hierarchy and no effects on the environment from the system, leading to no memory sources. However, for modes close to the cutoff and near the superhorizon regime, the characteristic timescales of the UV and IR modes become comparable. In this case, the backreaction of the UV modes on the IR modes cannot be neglected for such time scales, and the resulting effects are intrinsically non-Markovian.\footnote{{This type of non-Markovianity is distinct from that arising from colored noise induced by the use of a window function other than a step function.}} In fact, several works have investigated such non-Markovian effects~\cite{Cruces:2022imf,Figueroa:2020jkf,Figueroa:2021zah,Tomberg:2022mkt,Tomberg:2024evi,Brahma:2024ycc,Calderon-Figueroa:2025dto,Launay:2025kef,Launay:2025lnc,Kawasaki:2026hnx,Cruces:2026yvs}. Note that while such non-Markovianity is essential for the purpose of cosmological perturbations, it is not the case when considering the leading late-time contribution responsible for the IR divergence, which is {well-founded} motivation for {using} the stochastic formalism. In order to resum the leading order of secular growth, IR divergences, it is enough to consider a Markovian system since other contributions characterized by the much smaller time scale compared to the late time limit {are suppressed}. However, {if we want to use} the stochastic inflation {formalism} to describe non-linear cosmological perturbations, the non-Markovianity naturally arises because the finite time scale is considered like $N\sim \mathcal{O}(10)$ and all of time scale contributions could be of the same order. 
	\item[Quadratic (higher order) noises]\mbox{} \\
    In the heuristic picture of stochastic inflation, the IR modes are treated as a classical background, while the UV modes are treated as quantum linear perturbations. Since only linear UV modes are considered, the stochastic noise obeys Gaussian statistics, and terms quadratic (and, more generally, higher order) in the noise, corresponding to products of UV modes evaluated at the cutoff, are neglected. As far as we know, all existing works on stochastic inflation that aim to describe nonlinear perturbations employ this treatment. However, in this paper, we explicitly point out for the first time that, under the treatment of UV modes as linear perturbations, only linear perturbations of the IR modes are consistently justified. In other words, nonlinear perturbations of the IR modes cannot be consistently described without including terms quadratic in the noise. In fact, this statement was already implied in our previous work \cite{Cruces:2026yvs}.
\end{description}

In this work, we reformulate the heuristic Langevin description of the IR modes to consistently describe nonlinear perturbations, in particular, by incorporating both non-Markovian effects and terms quadratic in the noise. In \Sec{Sec:2}, we first review the standard stochastic formalism, in which terms quadratic in the noise are neglected, and show that this treatment is consistent at linear order. We then extend the Langevin-type description to the nonlinear regime by incorporating these contributions and compute the three-point correlators for a spectator field on a fixed de Sitter background and for a slow-rolling inflaton. In \Sec{Sec:3}, we compute the corresponding three-point correlators using the in-in formalism and explicitly demonstrate their agreement with those obtained from our reformulated stochastic framework, within the scope of superhorizon nonlinear evolution. Finally, in \Sec{Sec:4}, which presents the central message of this paper, we discuss the implications of these results for previous works based on assumptions that are not quantitatively justified for nonlinear perturbations, and reflect on the role and motivation of stochastic inflation in describing nonlinear cosmological perturbations of the inflaton.

\section{Reformulation of the stochastic inflation to the non-linear regime}
\label{Sec:2}

In this section, we present a formulation of stochastic inflation that consistently describes perturbations nonlinear in amplitude. To this end, we first review the conventional heuristic derivation of the stochastic formalism, which is based on cosmological intuition rather than on quantum field theory. However, as we show below, this derivation is valid only in the linear regime; namely, it cannot quantitatively describe perturbations nonlinear in amplitude. We therefore extend the heuristic derivation to the nonlinear regime in a perturbative way.

We consider a system consisting of a single scalar field $\phi$, whose action is given by
\begin{align}
S=\int \dd^4 x \sqrt{-g}\com{\frac{\mpl^2}{2}R-\frac{1}{2}g^{\mu\nu}\partial_\mu\phi\partial_\nu\phi - V(\phi)} . \label{action}
\end{align}
Here, $g^{\mu\nu}$ denotes the inverse metric, and $V(\phi)$ is the scalar potential. For later convenience, throughout this paper we use the number of~\efolds\ $N$, defined by $\dd N=H\dd t$, as the time variable.

In stochastic inflation, fields are decomposed into IR and UV modes. The separation is implemented by means of a window function $W(x)$, which, at leading order, is taken to be a step function, $W(x)\rightarrow\theta(1-x)$. For a generic quantity $Q$, the IR and UV components are defined as
\begin{align}
Q_\mathrm{IR} \equiv \int_\mathbf{k} e^{i \mathbf{k}\cdot\mathbf{x}} W\left(\frac{k}{k_\sigma(N)}\right) Q(N,\mathbf k) ,\qquad Q_\mathrm{UV} \equiv \int_\mathbf{k} e^{i \mathbf{k}\cdot\mathbf{x}} \com{1-W\left(\frac{k}{k_\sigma(N)}\right)} Q(N,\mathbf k) ,
\end{align}
where $\int_\mathbf{k}\equiv \int\frac{\dd\mathbf{k}}{(2\pi)^3}$. Here, the coarse-graining scale, or cutoff scale, is defined by
\begin{align}
k_\sigma(N) = \sigma a H_\mathrm{IR},
\end{align}
with $\sigma\ll 1$~\cite{Grain:2017dqa,Andersen:2021lii}.

Intuitively, the stochastic formalism proceeds as follows. We typically work in the spatially flat gauge, which is equivalent to the uniform-$N$ gauge on superhorizon scales under the separate-universe approximation~\cite{Sasaki:1998ug,Wands:2000dp,Lyth:2003im,Lyth:2004gb}\footnote{Throughout this paper, we adopt the definition of the separate-universe approximation used in Ref.~\cite{Cruces:2022imf}, namely the approximation in which both nonlocal terms and the momentum constraint of general relativity are neglected \cite{Tanaka:2021dww,Tanaka:2024mzw,Tanaka:2026zew}. As shown in Ref.~\cite{Cruces:2022imf}, when these contributions are retained, the spatially flat gauge and the uniform-$N$ gauge are no longer equivalent on superhorizon scales, although their difference is exponentially suppressed in most situations. A direct comparison between these two gauges on superhorizon scales, in regimes relevant to primordial black hole formation, can be found in Appendix~C of Ref.~\cite{Figueroa:2021zah}.} The UV modes are treated quantum mechanically, typically within perturbation theory specified by the Bunch--Davies vacuum. By contrast, the IR dynamics are effectively classical, and spatial-gradient terms can be neglected. As a result, each coarse-grained patch evolves independently within the separate-universe approximation. As UV modes cross the coarse-graining scale $k_\sigma(N)$, they are transferred to the IR sector, thereby inducing stochastic fluctuations that drive the IR background away from the homogeneous background solution. This continuous inflow of modes is modeled by stochastic noise terms, leading to Langevin equations governing the dynamics of the IR modes.

\subsection{Linear system based on a heuristic approach}\label{Sec:2linear}
Let us explicitly derive the EoMs for the IR modes, namely the Langevin equations, based on the intuitive picture discussed above, although their validity must ultimately be confirmed within a quantum field theoretical framework.

The Hamilton equations derived from the action for the inflaton \eqref{action} are given by
\begin{align}
    \dif{\phi}{N} &= {\pi} ,  \label{phi}\\
    \dif{\pi}{N} &= -\left(3-\frac{{\pi}^2}{2\mpl^2}\right)\pi - \frac{V_{,\phi}(\phi)}{H^2} + \frac{\nabla^2\phi}{a^2 H^2},\label{pi }
\end{align}
where $V_{,\phi}(\phi)$ denotes the derivative of the potential with respect to $\phi$. Note that, in our notation, the field and momentum variables have the same dimension. In the heuristic derivation, the decomposition $\phi = \phi_\uir + \phi_\uuv$ is substituted into the above EoMs. Keeping the leading order in the gradient expansion for the IR modes, the resulting equations are given, up to linear order in the amplitudes, by
\begin{align}
    \dif{\phi_\uir}{N} &= {\pi_\uir} + \Xi_\phi+ \int_\mathbf{k}e^{i\mathbf{k\cdot x}} \com{1 - W\prn{{\frac{k}{k_\sigma(N)}}} } E_\phi (N,\mathbf k) ,  \label{phi separation}\\
    \dif{\pi_\uir}{N} &= -\left(3-\frac{{\pi_\uir}^2}{2\mpl^2}\right)\pi_\uir - \frac{V_{,\phi}(\phi_\uir)}{H^2_\uir}  + \Xi_\pi  + \int_\mathbf{k} e^{i\mathbf{k\cdot x}} \com{1 - W\prn{{\frac{k}{k_\sigma(N)}}} }   E_\pi (N,\mathbf k),\label{pi separation}
\end{align}
where $\Xi_\phi$ and $\Xi_\pi$ are defined by
\begin{align}
    \Xi_\phi \equiv \int_\mathbf{k} e^{i\mathbf{k\cdot x}} \dif{W\prn{\frac{k}{k_\sigma}}}{N} \phi(N,\mathbf{k}),\quad
    \Xi_\pi \equiv \int_\mathbf{k} e^{i\mathbf{k\cdot x}} \dif{W\prn{\frac{k}{k_\sigma}}}{N} \pi(N,\mathbf{k}),
\end{align}
and the equations for the UV modes are expressed as
\begin{align}
     E_\phi(N,\mathbf k) &\equiv -\pdif{\phi(N,\mathbf k)}{N} + \pi(N,\mathbf k),\label{UV EoM phi}\\
     E_\pi(N,\mathbf k) &\equiv - \pdif{\pi(N,\mathbf k)}{N} - \frac{\partial}{\partial\pi_\uir}\left[\left(3-\frac{{\pi_\uir}^2}{2\mpl^2}\right)\pi_\uir + \frac{V_{,\phi}(\phi_\uir)}{H^2_\uir}\right]{\pi(N,\mathbf k) } \notag\\
     &\qquad- \frac{\partial}{\partial\phi_\uir}\left[\left(3-\frac{{\pi_\uir}^2}{2\mpl^2}\right)\pi_\uir + \frac{V_{,\phi}(\phi_\uir)}{H^2_\uir}\right]\phi(N,\mathbf k) - \frac{k^2}{a^2 H^2_\uir}\phi(N,\mathbf k). \label{UV EoM pi}
\end{align}

Let us emphasize that the equations above are originally derived as Heisenberg equations for quantum operators. In the language of the path integral, the classical tree-level EoMs for the IR modes are satisfied by the saddle-point configuration in the classical limit, while the UV modes are integrated out and affect the IR modes through the auxiliary fields $\Xi_\phi,\Xi_\pi$. To connect these equations with the intuitive picture of a classical IR background evolving under the influence of UV modes, we may interpret them as effective stochastic equations for ``classical'' variables, although the validity of this interpretation must ultimately be established within quantum theory.\footnote{\label{footnote:decoherence}Decoherence can support a classical statistical description of the IR modes by suppressing interference and establishing distinguishable records in an environment. However, in the Gaussian free theory, where the conventional stochastic formulation is valid (but requires the extensions developed in this paper beyond this regime), the UV modes do not induce decoherence of the IR modes~\cite{Habib:1992ci,Weenink:2011dd} and the UV environment does not record the IR configurations. Thus, the stochastic formulation should be regarded as an alternative computational framework rather than a description of objectively classical dynamics. One possible resolution is to introduce additional matter fields that decohere the IR modes and establish their classicality through the suppression of interference between alternatives, with the corresponding records formed in these fields.}\footnote{\label{footnote:CCB}Let us emphasize that the classicality'' associated with the decoherence of IR fields is conceptually distinct from, and never addresses, the emergence of one realized classical late-time universe, as is usually assumed in standard cosmology. Within the framework of Ref.~\cite{Kuroda2026CCB}, such a one-realization description requires the formation of \textit{Cosmological Classical Branches} (CCBs), characterized by a quantum state structure for matter components. (Squeezing and the suppression of interference alone do not establish a one-realization description beyond the statistical replacement of quantum correlators by classical averages.) For further clarification and a pedagogical discussion of ``classicality'' as a relative notion in quantum theory for cosmologists, see Refs.~\cite{Kuroda2026CCB,Kuroda:2026mtp} and the references therein.} 

Therefore, \textit{assuming} that the UV-mode operators satisfy $E_\phi = E_\pi = 0$, the dynamics of the classical IR configuration is governed by the Langevin equations
\begin{align}
    \dif{\phi_\uir}{N} &= {\pi_\uir} + \Xi_\phi , \label{phi IR}\\
    \dif{\pi_\uir}{N} &= -\left(3-\frac{{\pi_\uir}^2}{2\mpl^2}\right)\pi_\uir - \frac{V_{,\phi}(\phi_\uir)}{H^2_\uir} + \Xi_\pi . \label{pi IR}
\end{align}
Here, the noise terms $\Xi_\phi$ and $\Xi_\pi$ obey Gaussian statistics for a linear free field, with covariance\footnote{For simplicity, we focus on the dynamics of a single coarse-grained patch by invoking the separate-universe approximation. Consequently, the sinc function associated with correlations between patches at different spatial locations is neglected in the noise covariance.} given by
\begin{align}
    \Braket{\Xi_{X}(N)\Xi_{Y}(N')} &= \dif{\ln k_\sigma}{N}\mathcal{P}_{XY}(N,k_\sigma(N))\delta(N-N')\notag\\
    &= \left.\prn{\prn{1-\epsilon_{1,\uir}}\frac{k^3}{2\pi^2}(XY^{*})}\right|_{k=k_\sigma(N)} \delta(N-N'), \label{variance of noise}
\end{align}
where $X,Y\in\{\phi,\pi\}$, and the slow-roll parameter is defined by
\begin{align}
    \epsilon_{1,\uir} \equiv - \dif{\ln H_\uir}{N}.
\end{align}
The Hubble parameter is determined by the Friedmann equation,
\begin{align}
    3\mpl^2 H_\uir^2 = \frac{H_\uir^2\pi_\uir^2}{2} + V(\phi_\uir).
\end{align}
Formally taking the limit $\mpl\rightarrow\infty$ with $H=\mathrm{const}$ removes the dynamical effects of gravity and reproduces the case of a spectator field on a fixed de Sitter background.

Let us now discuss the regime of validity of this formulation. In conclusion, the above derivation can be quantitatively justified only for \textit{linear} perturbations in amplitude around the fiducial background, for both the IR and UV modes. First, as discussed above, the operator equations for the UV modes hold for a free linear field, which means that perturbativity in the self-couplings or gravitational couplings is assumed in the action and that only terms up to linear order in these couplings are taken into account in the derivation. Thus, at least for the UV modes, the action is treated perturbatively only up to linear order. On the other hand, it is sometimes argued in the literature that this restriction need not apply to the IR sector, and that the IR modes may therefore be nonlinear or even non-perturbative in amplitude. However, it is not the case because the regime of validity of the perturbative expansion, which is determined by the functional form of the action in terms of $\phi$ and its field value, is common to IR modes and UV modes. In other words, in the stochastic inflation, the one field governed by the unique action is merely decomposed into IR and UV components and thus the legitimate perturbative order coincides with $\phi,\phi_\uir,\phi_\uuv$. For instance, considering large values of the IR modes would, if anything, further worsen perturbative control. 

One concrete example of this perturbative limitation concerns the assumption $E_\phi=E_\pi=0$ made above. At linear order, this assumption is justified by the Mukhanov–Sasaki equations \eqref{phi UV first} and \eqref{pi UV first} for UV modes with $k>k_\sigma$, derived around a fixed homogeneous background. At higher orders, however, these linear equations are no longer sufficient, and the corresponding higher-order equations of motion for $k>k_\sigma$, such as Eq.~\eqref{quantum_EOM_second}, must instead be taken into account.

In this sense, predictions and claims made within the conventional approach must be understood as extrapolations of the linear description into the nonlinear, nonperturbative regime, providing at most speculation about possible qualitative features rather than quantitatively justified theoretical predictions. Such speculation acquires physical meaning only after being justified or confirmed by more fundamental methods valid in the nonlinear regime. More specifically, {if the Langevin equations used in conventional approaches are assumed to remain valid in the non-linear regime, one can mathematically get arbitrarily large perturbation amplitudes}. However, once the evolution goes beyond their regime of validity, namely that of linear perturbations, the resulting values, although mathematically well defined, cannot be regarded as quantitatively justified physical predictions. As explained above, the conventional Langevin equations are not quantitatively justified by the underlying physical theory in the nonlinear regime. In that regime, the governing equations themselves must instead be modified, as we discuss in the next subsection, Sec.~\ref{sec:nonlinear}.

Therefore, the system described by the Langevin equations \eqref{phi IR} and \eqref{pi IR} can equivalently be rewritten as follows. Let us assume that the amplitudes of the noise terms are much smaller than the IR quantities, as is indeed the case during inflation, and count them as being of the same perturbative order as the perturbations in $\phi_\uir$ and $\pi_\uir$, respectively. Then, the perturbative expansion $\phi_\uir = \bar\phi + \phi^{(1)}_\uir,\ \pi_\uir = \bar\pi + \pi^{(1)}_\uir,\ \Xi= \Xi^{(1)}$ yields the following system:
\begin{description}
    \item[0th order (homogeneous background)]\mbox{} \\
    At this order, the equations simply reduce to the homogeneous background equations:
\begin{align}
    \dif{\bar\phi}{N} &= \bar{\pi}, \label{phi background}\\
    \dif{\bar\pi}{N} &= -\left(3-\frac{{\bar\pi}^2}{2\mpl^2}\right)\bar\pi - \frac{V_{,\phi}(\bar\phi)}{\bar H^2}.
    \label{pi background}
\end{align} \label{0theq SNE}

    \item[1st order]\mbox{} \\
The first-order equations for the IR modes are given by
\begin{align}
    \dif{\phi_\uir^{(1)}}{N} &= {\pi_\uir^{(1)}} + \Xi_\phi^{(1)},  \label{phi IR first}\\
    \dif{\pi_\uir^{(1)}}{N} &= - \frac{\partial}{\partial\bar\pi}\left[\left(3-\frac{{\bar\pi}^2}{2\mpl^2}\right)\bar\pi + \frac{V_{,\phi}(\bar\phi)}{\bar H^2}\right]{\pi_\uir^{(1)} } - \frac{\partial}{\partial\bar\phi}\left[\left(3-\frac{{\bar\pi}^2}{2\mpl^2}\right)\bar\pi + \frac{V_{,\phi}(\bar\phi)}{\bar H^2}\right]\phi_\uir^{(1)} + \Xi_\pi^{(1)},\label{pi IR first} 
\end{align}
or, equivalently,
\begin{align}
     {F}\bar{\pi}\dif{}{N}\com{\frac{1}{{F}}\dif{}{N}\left(\frac{\phi^{(1)}_\uir}{\bar{\pi}}\right)}=\dif{\Xi_\phi^{(1)}}{N}+3\Xi_\phi^{(1)}+\Xi_\pi^{(1)},
    \label{first_IR_modified}
\end{align}
where
\begin{align}
    F(N)\equiv 2\mpl^2\frac{3-\frac{\bar{\pi}^2(N)}{2\mpl^2}}{3a^3(N)\bar{\pi}^2(N) \bar H(\bar{\phi}(N),\bar{\pi}(N))} {\simeq \frac{2\mpl^2}{a^3(N)\bar{\pi}^2(N) \bar H(\bar{\phi}(N),\bar{\pi}(N))}},
\end{align}
{where we neglect the decaying mode for consistency with the separate-universe approximation~\cite{Cruces:2022imf}.\footnote{The difference between \Eq{pi IR first} and \Eq{pi UV first} in the $k\rightarrow 0$ limit can also be neglected within the separate-universe approximation.}}
The UV modes follow the Mukhanov--Sasaki equations on the deterministic background for $k>k_\sigma$,
\begin{align}
    0&=\pdif{\phi^{(1)}(N,\mathbf{k})}{N} - \pi^{(1)}(N,\mathbf{k}) ,  \label{phi UV first}\\
    0&= \pdif{\pi^{(1)}}{N} + \left(3-\frac{{\bar\pi}^2}{2\mpl^2}\right)\pi^{(1)} + \com{\frac{V_{,\phi\phi}(\bar\phi)}{\bar H^2} - \frac{1}{a^3\mpl^2 \bar H}\dif{}{N}\prn{a^3 \bar H \bar \pi^2}} \phi^{(1)} + \frac{k^2}{a^2\bar H^2} \phi^{(1)}\label{pi UV first}
\end{align}
or, equivalently,
\begin{align}
     {F}\bar{\pi}\dif{}{N}\com{\frac{1}{{F}}\dif{}{N}\left(\frac{\phi^{(1)}(N,\mathbf{k})}{\bar{\pi}}\right)}=-\frac{k^2}{a^2\bar H^2} \phi^{(1)}(N,\mathbf{k}),
    \label{first_UV_modified}
\end{align}
together with the Bunch--Davies vacuum condition in the infinite past.

These equations can be solved formally, yielding\footnote{\label{foot:real_pert}{Note that $\phi^{(1)}$ is evaluated only on superhorizon scales in the integrals. Since its dominant superhorizon contribution can be made real by an appropriate choice of the phase of each mode function with the Bunch--Davies vacuum, we take $\phi^{(1)}$ to be real both in these integrals and throughout the following discussion.}} 
\begin{align}   
\phi_\uir^{(1)}=\int_0^{N}\dd N_1\sqrt{\frac{\left(\sigma a(N_1)H\right)^3}{2\pi^2}}\phi^{(1)}\left(N,\sigma a(N_1) H\right)\xi(N_1) + \varphi_\mathrm{hom}[\phi_\uir^{(1)}(0);N]
\label{first_IR_solution}
\end{align}
where the normalized noise $\xi$ is introduced with the covariance
\begin{equation}
\langle\xi(N_1)\xi(N_2)\rangle=\delta(N_1-N_2).
\end{equation}
Here the initial conditions at $N=0$ are encoded in the homogeneous solution of Eq.~\eqref{first_IR_modified}:
\begin{align}
    \varphi_\mathrm{hom}[X(N');N] \equiv \bar\pi(N)\com{\frac{X(N')}{\bar\pi(N')}  + \frac{1}{F(N')} \dif{}{N}\prn{\frac{X}{\bar\pi}}\Bigg|_{N=N'} \int_{N'}^N \dd N_1 F(N_1) } \label{def of hom}
\end{align}
for any function $X(N)$. Indeed, using this solution, the one-point function is given by
\begin{align}
    \Braket{\phi_\uir^{(1)}(N)}
    =
    \Braket{\varphi_\mathrm{hom}[\phi_\uir^{(1)}(0);N]}.
\end{align}
This contribution is purely homogeneous and can therefore be absorbed into a redefinition of the homogeneous background by $\bar\phi(N)\rightarrow\bar\phi(N)+\varphi_{\rm hom}[\phi_{\rm IR}^{(1)}(0);N]$. After this background shift, the corresponding perturbation has a vanishing one-point function. In what follows, we adopt this redefined background and denote the perturbation around it again by $\phi_\uir^{(1)}$. The linear perturbation then obeys Gaussian statistics with
\begin{align}
    \Braket{\phi_\uir^{(1)}(N)}=0.
\end{align}
Consequently, the two-point correlation function, which is the only independent statistical quantity at linear order, is correctly reproduced as
\begin{align}
    \Braket{\phi_\uir^{(1)}(N)^2}&=\int_0^N\dd N_1\frac{\left(\sigma a(N_1)H\right)^3}{2\pi^2}\left|\phi^{(1)}\left(N,\sigma a(N_1) H\right)\right|^2\notag\\
    &= \int_{\sigma a(0) H}^{\sigma a(N) H}\frac{\dd k}{k}\frac{k^3}{2\pi^2}\left|\phi^{(1)}\left(N,k\right)\right|^2\notag \\
    &{\simeq \int_{\sigma a(0) H}^{\sigma a(N) H}\frac{\dd k}{k}\frac{k^3}{2\pi^2}P_k(N,N)}
    \label{first_2-point}
\end{align}
{where we define $P_k(N,N_1)\equiv \operatorname{Re}\left[ \delta\phi(N)\delta\phi^{*}(N_1)\right]$  and in this case $P_k(N,N)\simeq \left|\phi^{(1)}\left(N,k\right)\right|^2$ (see footnote \ref{foot:real_pert}).}

\end{description}

\subsection{Perturbative extension to the nonlinear regime}
\label{sec:nonlinear}
To describe \textit{nonlinear} perturbations, the derivation presented in the previous subsection must be modified, as we demonstrate below. As a first step toward this extension, we consider perturbations up to second order. At nonlinear order, it is essential to account for the momentum carried by products of perturbations, since the Fourier projection of a product is not, in general, equivalent to the product of separately projected fields. This distinction gives rise to additional contributions neglected in the conventional approach, including non-Markovian effects and noise terms quadratic in UV fluctuations, which are essential for quantitative agreement with quantum-field-theoretic predictions.

\subsubsection{Dynamical equations for IR modes at second order}\label{sec:nonlinear1}

For simplicity, let us first consider a spectator field on top of the fixed de Sitter spacetime and discuss the case of the inflaton including dynamical gravity in the following subsection \ref{sec:nonlinear3}. Following the same procedure as in the previous subsection, we consider the Heisenberg equations for quantum operators. Formally expanding the equations up to second order, the EoM becomes
\begin{equation}
    \dif{^2\phi^{(2)}}{N^2}+3\dif{\phi^{(2)}}{N}+\frac{V_{\phi\phi}(\bar{\phi})}{H^2}\phi^{(2)}-\frac{\nabla^2}{a^2 H^2}\phi^{(2)}+\frac{1}{2}\frac{V_{\phi\phi\phi}(\bar{\phi})}{H^2}\left(\phi^{(1)}\right)^2=0,
    \label{LPT_second}
\end{equation}
or equivalently
\begin{align}
    F\bar\pi\dif{}{N}\left[\frac{1}{F}\dif{}{N}\left(\frac{\phi^{(2)}}{\bar{\pi}}\right)\right]= -\frac{1}{2}\frac{V_{\phi\phi\phi}(\bar\phi)}{H^2}\left(\phi^{(1)}\right)^2 +\frac{\nabla^2}{a^2 H^2}\phi^{(2)}.
    \label{Second order EoM deSitter}
\end{align}
Splitting the field into IR and UV modes $\phi=\phi_\uir^{(2)} + \phi_\uuv^{(2)}$ then yields the following exact projected equation:
\begin{small}
\begin{align}
    &\dif{^2\phi^{(2)}_\uir}{N^2}+3\dif{\phi^{(2)}_\uir}{N}
    +\frac{V_{\phi\phi}(\bar\phi)}{H^2}\phi^{(2)}_\uir
    +\int_{\mathbf k}e^{i\mathbf k\cdot\mathbf x}
    W\prn{\frac{k}{k_\sigma(N)}}\frac{k^2}{a^2H^2}\phi^{(2)}(N,\mathbf k)\notag\\
    &\quad+\frac12\frac{V_{\phi\phi\phi}(\bar\phi)}{H^2}
    \int_{\mathbf k}e^{i\mathbf k\cdot\mathbf x}W\prn{\frac{k}{k_\sigma(N)}}
    \int_{\mathbf q}\phi^{(1)}(N,\mathbf q)\phi^{(1)}(N,\mathbf k-\mathbf q)
    =\dif{\Xi_\phi^{(2)}}{N}+3\Xi_\phi^{(2)}+\Xi_\pi^{(2)}\notag\\
    &-\int_\mathbf{k} e^{i\mathbf{k}\cdot\mathbf{x}} \com{1 - W\prn{{\frac{k}{k_\sigma(N)}}} } \Bigg[\dif{^2{\phi}^{(2)}(N,\mathbf{k})}{N^2}+3\dif{{\phi}^{(2)}(N,\mathbf{k})}{N}+\frac{V_{\phi\phi}(\bar{\phi})}{H^2}{\phi}^{(2)}(N,\mathbf{k})+\frac{k^2}{a^2H^2}{\phi}^{(2)}(N,\mathbf{k})\notag
    \\&+\frac{1}{2}\frac{V_{\phi\phi\phi}(\bar{\phi})}{H^2}\int_\mathbf{q} {\phi}^{(1)}(N,\mathbf{q}) {\phi}^{(1)}(N,\mathbf{k}-\mathbf{q})\Bigg]
    \label{full_stochastic_second}
\end{align}
\end{small}
where the definitions of the noise terms are given by
\begin{align}
    \Xi_{\phi}^{(2)}&\equiv\int_{\mathbf{k}}e^{i\mathbf{k}\cdot\mathbf{x}}
    \dif{W\prn{\frac{k}{k_\sigma(N)}}}{N}\phi^{(2)}(N,\mathbf{k}),\quad\Xi_{\pi}^{(2)}\equiv\int_{\mathbf{k}}e^{i\mathbf{k}\cdot\mathbf{x}}
    \dif{W\prn{\frac{k}{k_\sigma(N)}}}{N}
    \dif{\phi^{(2)}(N,\mathbf{k})}{N}.
    \label{quantum_noises_second}
\end{align}
The last two lines in \eqref{full_stochastic_second} actually vanish due to the Fourier expression of Eq.~\eqref{LPT_second} as the underlying second-order UV-mode operator for \(k>k_\sigma(N)\):
\begin{align}
    &\dif{^2{\hat\phi}^{(2)}(N,\mathbf{k})}{N^2}+3\dif{{\hat\phi}^{(2)}(N,\mathbf{k})}{N}+\frac{V_{\phi\phi}(\bar{\phi})}{H^2}{\hat\phi}^{(2)}(N,\mathbf{k})+\frac{k^2}{a^2H^2}{\hat\phi}^{(2)}(N,\mathbf{k})\notag
    \\&+\frac{1}{2}\frac{V_{\phi\phi\phi}(\bar{\phi})}{H^2}\int_\mathbf{q} {\hat\phi}^{(1)}(N,\mathbf{q}) {\hat\phi}^{(1)}(N,\mathbf{k}-\mathbf{q})=0. \label{quantum_EOM_second}
\end{align}
For a step-function window, the stochastic noise is determined by the UV modes evaluated at the moving coarse-graining scale $k=k_\sigma(N)$, so their nonlinear evolution must also be taken into account. Importantly, two first-order IR modes can source a second-order UV mode, while two first-order UV modes can source a second-order IR mode, since the IR/UV classification of the output momentum is not determined solely by those of the individual input momenta. Quadratic interactions must therefore be retained in both the projected IR equation and the UV equations governing the modes that enter the noise. In particular, the last term in Eq.~\eqref{quantum_EOM_second} involves both IR and UV components of the internal momenta, even when the external momentum $\mathbf{k}$ lies in the UV sector. As shown below, these interactions give rise to contributions neglected in the conventional approach, including non-Markovian effects associated with $\phi_\uir\phi_\uuv$ and quadratic noise terms arising from $\phi_\uuv\phi_\uuv$. A comparison with tree-level quantum field theory from the perspective of the equations of motion is presented in Appendix~\ref{sec:tree-initial-value}.

Therefore, neglecting the IR spatial-gradient term in Eq.~\eqref{full_stochastic_second}, the second-order IR dynamical equations become
\begin{align}
    &\dif{^2\phi^{(2)}_\uir}{N^2}+3\dif{\phi^{(2)}_\uir}{N}
    +\frac{V_{\phi\phi}(\bar\phi)}{H^2}\phi^{(2)}_\uir+\frac12\frac{V_{\phi\phi\phi}(\bar\phi)}{H^2}
    \int_{\mathbf k}e^{i\mathbf k\cdot\mathbf x}W\prn{\frac{k}{k_\sigma(N)}}
    \int_{\mathbf q}\phi^{(1)}(N,\mathbf q)\phi^{(1)}(N,\mathbf k-\mathbf q)\notag\\
    &\hspace{100mm}
    =\dif{\Xi_\phi^{(2)}}{N}+3\Xi_\phi^{(2)}+\Xi_\pi^{(2)}
    \label{second_IR}
\end{align}
or, equivalently,
\begin{align}
    &F\bar\pi\dif{}{N}\left[\frac1F\dif{}{N}\left(\frac{\phi_\uir^{(2)}}{\bar\pi}\right)\right]\notag\\
    &=-\frac12\frac{V_{\phi\phi\phi}}{H^2}
    \int_{\mathbf k}e^{i\mathbf k\cdot\mathbf x}W\prn{\frac{k}{k_\sigma(N)}}
    \int_{\mathbf q}\phi^{(1)}(N,\mathbf q)\phi^{(1)}(N,\mathbf k-\mathbf q)
    +\dif{\Xi_\phi^{(2)}}{N}+3\Xi_\phi^{(2)}+\Xi_\pi^{(2)}.
    \label{second_IR_modified}
\end{align}

\subsubsection{Three-point correlators for superhorizon non-linear evolution}\label{sec:nonlinear2}

In what follows, we compute the three-point function arising from nonlinear evolution on superhorizon scales. In particular, nonlinear interactions at horizon crossing and on subhorizon scales are neglected by construction. To evaluate the correlator at time $N$, we adopt the following strategy. We neglect all terms proportional to $k^2$ in the equations of motion for both the IR and UV modes and consider the resulting equations over the interval from $N_{\mathrm{s.h.}}(N)$ to $N$, where $N_{\mathrm{s.h.}}(N)$ denotes a time shortly after horizon crossing of the mode with $k=\sigma a(N)H(N)$, at which the superhorizon approximation becomes valid. As we will show, these equations can be solved explicitly by imposing initial conditions that eliminate nonlinear contributions generated at horizon crossing and on subhorizon scales. We then repeat the same procedure over the preceding interval, from $N_{\mathrm{s.h.}}(N_{\mathrm{s.h.}}(N))$ to $N_{\mathrm{s.h.}}(N)$, and successively over earlier intervals. By combining the solutions obtained for these successive intervals, we can reconstruct the nonlinear superhorizon evolution from $N=0$ to $N$, where $N=0$ corresponds to the IR cutoff of interest.

To make the computation transparent, we retain the decomposition of Eq.~\eqref{second_IR_modified} into three contributions, labeled \(\mathrm a\) (IR--IR), \(\mathrm b\) (IR--UV), and \(\mathrm c\) (UV--UV). In particular, recalling the IR and UV parts of a linear perturbation in Fourier space, as defined above, given by
\begin{align}
    \phi_{\uir}^{(1)}(N,\mathbf{k})&=W\prn{\frac{k}{k_\sigma(N)}}\phi^{(1)}(N,\mathbf{k}),\quad 
    \phi_{\uuv}^{(1)}(N,\mathbf{k})=\com{1-W\prn{\frac{k}{k_\sigma(N)}}}\phi^{(1)}(N,\mathbf{k}),
    \label{IR UV Fourier second order note}
\end{align}
the nonlinear convolution term is decomposed as
\begin{align}
    &\int_{\mathbf q}\phi^{(1)}(N,\mathbf q)\phi^{(1)}(N,\mathbf k-\mathbf q)\notag\\
    &\quad=\int_{\mathbf q}\Big[\phi_{\uir}^{(1)}(N,\mathbf q)\phi_{\uir}^{(1)}(N,\mathbf k-\mathbf q)
    +2\phi_{\uir}^{(1)}(N,\mathbf q)\phi_{\uuv}^{(1)}(N,\mathbf k-\mathbf q)
    +\phi_{\uuv}^{(1)}(N,\mathbf q)\phi_{\uuv}^{(1)}(N,\mathbf k-\mathbf q)\Big].
    \label{quadratic Fourier split}
\end{align}
Because this nonlinear source can generate either an IR or a UV output momentum, each contribution of Eq.~\eqref{second_IR_modified} now has both a directly projected source and a possible second-order noise term:
\begin{small}
\begin{align}
    &F\bar\pi\dif{}{N}\left[\frac1F\dif{}{N}\left(\frac{\phi_\uir^{(2)\mathrm a}}{\bar\pi}\right)\right]
    =-\frac12\frac{V_{\phi\phi\phi}}{H^2}\int_{\mathbf k}e^{i\mathbf k\cdot\mathbf x}W\prn{\frac{k}{k_\sigma(N)}}
    \int_{\mathbf q}\phi_\uir^{(1)}(N,\mathbf q)\phi_\uir^{(1)}(N,\mathbf k-\mathbf q)
    +\dif{\Xi_\phi^{(2)\mathrm a}}{N}+3\Xi_\phi^{(2)\mathrm a}+\Xi_\pi^{(2)\mathrm a},
    \label{second_IR_split1}\\
    &F\bar\pi\dif{}{N}\left[\frac1F\dif{}{N}\left(\frac{\phi_\uir^{(2)\mathrm b}}{\bar\pi}\right)\right]
    =-\frac{V_{\phi\phi\phi}}{H^2}\int_{\mathbf k}e^{i\mathbf k\cdot\mathbf x}W\prn{\frac{k}{k_\sigma(N)}}
    \int_{\mathbf q}\phi_\uir^{(1)}(N,\mathbf q)\phi_\uuv^{(1)}(N,\mathbf k-\mathbf q)
    +\dif{\Xi_\phi^{(2)\mathrm b}}{N}+3\Xi_\phi^{(2)\mathrm b}+\Xi_\pi^{(2)\mathrm b},
    \label{second_IR_split2}\\
    &F\bar\pi\dif{}{N}\left[\frac1F\dif{}{N}\left(\frac{\phi_\uir^{(2)\mathrm c}}{\bar\pi}\right)\right]
    =-\frac12\frac{V_{\phi\phi\phi}}{H^2}\int_{\mathbf k}e^{i\mathbf k\cdot\mathbf x}W\prn{\frac{k}{k_\sigma(N)}}
    \int_{\mathbf q}\phi_\uuv^{(1)}(N,\mathbf q)\phi_\uuv^{(1)}(N,\mathbf k-\mathbf q)
    +\dif{\Xi_\phi^{(2)\mathrm c}}{N}+3\Xi_\phi^{(2)\mathrm c}+\Xi_\pi^{(2)\mathrm c}.
    \label{second_IR_split3}
\end{align}
\end{small}
For each \(i=\mathrm a,\mathrm b,\mathrm c\), the noise terms are defined, in the notation of Eq.~\eqref{quantum_noises_second}, by
\begin{align}
    \Xi_\phi^{(2)i}(N,\mathbf x)&\equiv\int_{\mathbf k}e^{i\mathbf k\cdot\mathbf x}
    \dif{W\prn{\frac{k}{k_\sigma(N)}}}{N}\phi^{(2)i}(N,\mathbf k),\quad
    \Xi_\pi^{(2)i}(N,\mathbf x)\equiv\int_{\mathbf k}e^{i\mathbf k\cdot\mathbf x}
    \dif{W\prn{\frac{k}{k_\sigma(N)}}}{N}\dif{\phi^{(2)i}(N,\mathbf k)}{N}.
    \label{quantum_noises_second bc}
\end{align}
The corresponding UV mode equation \eqref{quantum_EOM_second}, for \(k>k_\sigma(N)\), is also decomposed into 
\begin{align}
    F\bar\pi\dif{}{N}\left[\frac1F\dif{}{N}\left(\frac{\phi^{(2)\mathrm a}(N,\mathbf k)}{\bar\pi}\right)\right]
    &=-\frac12\frac{V_{\phi\phi\phi}}{H^2}\int_{\mathbf q}\phi_\uir^{(1)}(N,\mathbf q)\phi_\uir^{(1)}(N,\mathbf k-\mathbf q),
    \label{quantum_EOMS_second split a 2}\\
    F\bar\pi\dif{}{N}\left[\frac1F\dif{}{N}\left(\frac{\phi^{(2)\mathrm b}(N,\mathbf k)}{\bar\pi}\right)\right]
    &=-\frac{V_{\phi\phi\phi}}{H^2}\int_{\mathbf q}\phi_\uir^{(1)}(N,\mathbf q)\phi_\uuv^{(1)}(N,\mathbf k-\mathbf q),
    \label{quantum_EOMS_second split b 2}\\
    F\bar\pi\dif{}{N}\left[\frac1F\dif{}{N}\left(\frac{\phi^{(2)\mathrm c}(N,\mathbf k)}{\bar\pi}\right)\right]
    &=-\frac12\frac{V_{\phi\phi\phi}}{H^2}\int_{\mathbf q}\phi_\uuv^{(1)}(N,\mathbf q)\phi_\uuv^{(1)}(N,\mathbf k-\mathbf q),
    \label{quantum_EOMS_second split c 2}
\end{align}
sourcing each noise \eqref{quantum_noises_second bc}. Here and in the subsequent UV equations, we omit the gradient term proportional to $k^2/(a^2H^2)$ to focus on superhorizon nonlinear evolution. The label “a” denotes contributions arising solely from IR–IR interactions, while “b” denotes those from IR–UV interactions, which have already been considered in the literature and may give rise to non-Markovian effects. Finally, “c” denotes contributions from UV–UV interactions, which are quadratic in the UV fluctuations and are incorporated into the stochastic formulation for the first time in this work.

As discussed above, we focus on the superhorizon evolution of $\phi_\mathrm{UV}$ from $N_{\mathrm{s.h.}}(N)$ to $N$, during which the mode lies in the regime $\sigma aH < k \ll aH$. In this regime, the $k^2$ term can be neglected, and the $k$-dependence of $\phi_\mathrm{UV}$ enters only through the initial conditions. In this case, the solutions to Eqs.~\eqref{quantum_EOMS_second split a 2}--\eqref{quantum_EOMS_second split c 2} are given by 
\begin{align}
    &\phi^{(2)\mathrm a}(N,\mathbf{k})=\varphi_\mathrm{hom}[\phi^{(2)\mathrm a}(N_\mathrm{s.h.},\mathbf{k});N]\notag\\
    &\quad-\frac{\bar\pi(N)}2\int_{N_\mathrm{s.h.}}^N\dd N_1F(N_1)
    \int_{N_\mathrm{s.h.}}^{N_1}\dd N_2\frac{V_{\phi\phi\phi}(\bar\phi(N_2))}{F(N_2)\bar\pi(N_2)H^2}
    \int_{\mathbf q}\phi_\uir^{(1)}(N_2,\mathbf q)\phi_\uir^{(1)}(N_2,\mathbf k-\mathbf q),
    \label{second_UV_solution a}\\
    &\phi^{(2)\mathrm b}(N,\mathbf{k})=\varphi_\mathrm{hom}[\phi^{(2)\mathrm b}(N_\mathrm{s.h.},\mathbf{k});N]\notag\\
    &\quad-\bar\pi(N)\int_{N_\mathrm{s.h.}}^N\dd N_1F(N_1)
    \int_{N_\mathrm{s.h.}}^{N_1}\dd N_2\frac{V_{\phi\phi\phi}(\bar\phi(N_2))}{F(N_2)\bar\pi(N_2)H^2}
    \int_{\mathbf q}\phi_\uir^{(1)}(N_2,\mathbf q)\phi_\uuv^{(1)}(N_2,\mathbf k-\mathbf q),
    \label{second_UV_solution b}\\
    &\phi^{(2)\mathrm c}(N,\mathbf{k})=\varphi_\mathrm{hom}[\phi^{(2)\mathrm c}(N_\mathrm{s.h.},\mathbf{k});N]\notag\\
    &\quad-\frac{\bar\pi(N)}2\int_{N_\mathrm{s.h.}}^N\dd N_1F(N_1)
    \int_{N_\mathrm{s.h.}}^{N_1}\dd N_2\frac{V_{\phi\phi\phi}(\bar\phi(N_2))}{F(N_2)\bar\pi(N_2)H^2}
    \int_{\mathbf q}\phi_\uuv^{(1)}(N_2,\mathbf q)\phi_\uuv^{(1)}(N_2,\mathbf k-\mathbf q).
    \label{second_UV_solution c}
\end{align}
These formulas hold for the UV output modes at the respective time; a second-order mode generated outside the IR band can subsequently enter it through the noise in Eq.~\eqref{quantum_noises_second bc}. As will be discussed around Eq.~\eqref{initial condition UV2}, contributions from the initial conditions of the second-order modes (and then those of noises) are discarded when isolating the nonlinear evolution generated after \(N_\mathrm{s.h.}\).

\paragraph{Contribution labeled ``\(\mathrm{a}\)'': IR--IR interactions} \mbox{}\\
We begin by solving Eq.~\eqref{second_IR_split1} through integration over $N$. The resulting solution can be decomposed into three contributions:
\begin{align}
    \phi_\uir^{(2)\mathrm{a}} = \varphi_\mathrm{hom}[\phi_\uir^{(2)\mathrm a}(N_\mathrm{s.h.});N] + \left.\phi_\uir^{(2)\mathrm a} \right|_{\mathrm{direct}} + \left.\phi_\uir^{(2)\mathrm a} \right|_{\mathrm{noise}},
\end{align}
where the first term represents the homogeneous solution determined by the initial conditions, while the remaining two terms arise from the direct interaction and noise sources on the right-hand side of Eq.~\eqref{second_IR_split1}, respectively. These contributions are given by:
\begin{align}
    \left.\phi_\uir^{(2)\mathrm a}(N,\mathbf x)
    \right|_{\mathrm{direct}}
    &\equiv-\frac{\bar\pi(N)}{2}
    \int_{N_\mathrm{s.h.}}^N\dd N_1 F(N_1)
    \int_{N_\mathrm{s.h.}}^{N_1}\dd N_2
    \frac{V_{\phi\phi\phi}(\bar\phi(N_2))}
    {F(N_2)\bar\pi(N_2)H^2}\notag\\
    &\qquad\times\int_{\mathbf k}e^{i\mathbf k\cdot\mathbf x}
    W\prn{\frac{k}{k_\sigma(N_2)}}
    \int_{\mathbf q}
    \phi_\uir^{(1)}(N_2,\mathbf q)
    \phi_\uir^{(1)}(N_2,\mathbf k-\mathbf q).\label{second order solution direct}
\end{align}
\begin{align}
    \left.\phi_\uir^{(2)\mathrm a}(N,\mathbf x)
    \right|_{\mathrm{noise}}
    &\equiv\bar\pi(N)
    \int_{N_\mathrm{s.h.}}^N\dd N_1 F(N_1)
    \int_{N_\mathrm{s.h.}}^{N_1}\dd N_3
    \notag\\
    &\quad\times\frac{1}{F(N_3)\bar\pi(N_3)}
    \Bigg[
    \dif{\Xi_\phi^{(2)\mathrm a}(N_3,\mathbf x)}{N_3}
    +3\Xi_\phi^{(2)\mathrm a}(N_3,\mathbf x)
    +\Xi_\pi^{(2)\mathrm a}(N_3,\mathbf x)
    \Bigg]. \label{second order solution noise}
\end{align}

We first simplify the noise contribution before combining it with the direct contribution. Exchanging the order of integration, performing an assuming gaussian on the first term in Eq.~\eqref{second order solution noise}, and using
$\dif{}{N}\prn{\frac{1}{F\bar\pi}}=\frac{1}{F\bar\pi}\prn{3+\frac{1}{\bar\pi}\dif{\bar\pi}{N}}$
together with the definitions of the noise terms in Eq.~\eqref{quantum_noises_second bc}, we obtain
\begin{align}
    \left.\phi_\uir^{(2)\mathrm a}(N,\mathbf x)
    \right|_{\mathrm{noise}}
    &=\bar\pi(N)
    \int_{\mathbf k}e^{i\mathbf k\cdot\mathbf x}
    \int_{N_\mathrm{s.h.}}^N\dd N_3
    \dif{W\prn{\frac{k}{k_\sigma(N_3)}}}{N_3}
    \notag\\
    &\quad\times\Bigg[
    \frac{\phi^{(2)\mathrm a}(N_3,\mathbf k)}
    {\bar\pi(N_3)}
    +\frac{1}{F(N_3)}
    \int_{N_3}^N\dd N_1F(N_1)
    \dif{}{N_3}
    \prn{\frac{\phi^{(2)\mathrm a}(N_3,\mathbf k)}
    {\bar\pi(N_3)}}
    \Bigg].
\end{align}
Here, we neglect contributions associated with the initial UV modes and initial-time noise terms. Substituting Eq.~\eqref{second_UV_solution a} then yields
\begin{align}
    \left.\phi_\uir^{(2)\mathrm a}(N,\mathbf x)
    \right|_{\mathrm{noise}}
    &=-\frac{\bar\pi(N)}{2}
    \int_{N_\mathrm{s.h.}}^N\dd N_1 F(N_1)
    \int_{N_\mathrm{s.h.}}^{N_1}\dd N_2
    \frac{V_{\phi\phi\phi}(\bar\phi(N_2))}
    {F(N_2)\bar\pi(N_2)H^2}
    \notag\\
    &\quad\times
    \int_{\mathbf k}e^{i\mathbf k\cdot\mathbf x}
    \left[
    \int_{N_2}^N\dd N_3
    \dif{W\prn{\frac{k}{k_\sigma(N_3)}}}{N_3}
    \right]
    \int_{\mathbf q}
    \phi_\uir^{(1)}(N_2,\mathbf q)
    \phi_\uir^{(1)}(N_2,\mathbf k-\mathbf q),
    \label{second order solution noise 3}
\end{align}
Combining this noise contribution with the direct contribution in Eq.~\eqref{second order solution direct}, we find that their output-window factors satisfy
$W\prn{\frac{k}{k_\sigma(N_2)}}+\int_{N_2}^N\dd N_3\dif{W\prn{\frac{k}{k_\sigma(N_3)}}}{N_3}=W\prn{\frac{k}{k_\sigma(N)}}$.
The first term accounts for modes already in the IR sector at the interaction time $N_2$, while the second accounts for modes that subsequently enter the IR sector through the second-order noise. Thus, including the homogeneous contribution, we finally obtain
\begin{align}
    \phi_\uir^{(2)\mathrm a}(N,\mathbf x)
    &=\varphi_\mathrm{hom}
    [\phi_\uir^{(2)\mathrm a}(N_\mathrm{s.h.});N]-\frac{\bar\pi(N)}{2}
    \int_{N_\mathrm{s.h.}}^N\dd N_1 F(N_1)
    \int_{N_\mathrm{s.h.}}^{N_1}\dd N_2
    \frac{V_{\phi\phi\phi}(\bar\phi(N_2))}
    {F(N_2)\bar\pi(N_2)H^2}
    \notag\\
    &\hspace{30mm}\times
    \int_{\mathbf k}e^{i\mathbf k\cdot\mathbf x}
    W\prn{\frac{k}{k_\sigma(N)}}
    \int_{\mathbf q}
    \phi_\uir^{(1)}(N_2,\mathbf q)
    \phi_\uir^{(1)}(N_2,\mathbf k-\mathbf q).
    \label{second_IR1_solution}
\end{align}

We now evaluate the connected three-point function at tree level. Since $\phi_\uir^{(1)}$ is Gaussian, the leading nonvanishing contribution involves one second-order and two first-order perturbations:
\begin{align}
   \langle \left(\phi_\uir(N)-\langle\phi_\uir(N)\rangle\right)^3\rangle
   &\supset \left.\langle \left(\phi_\uir(N)-\langle\phi_\uir(N)\rangle\right)^3\rangle\right|_\mathrm{a}\notag\\
   &\equiv 3\left(\langle \phi_\uir^{(1)}(N)^2 \phi^{(2)\mathrm{a}}_\uir(N)\rangle-\langle\phi^{(1)}_\uir(N)^2\rangle\langle\phi_\uir^{(2)\mathrm{a}}(N)\rangle\right) + \mathcal{O}(\phi^{(6)}).
   \label{3-point_definition}
\end{align}
Here, $\langle\phi^{(1)}_\uir(N)\rangle=0$, the variance $\langle\phi^{(1)}_\uir(N)^2\rangle$ is given by Eq.~\eqref{first_2-point}, and $\mathcal{O}(\phi^{(n)})$ denotes the total perturbative order in the fluctuation amplitude. We therefore evaluate $\langle\phi_\uir^{(2)\mathrm{a}}(N)\rangle$ and $\langle \phi_\uir^{(1)}(N)^2 \phi^{(2)\mathrm{a}}_\uir(N)\rangle$ separately. Substituting the linear-order solution \eqref{first_IR_solution} into Eq.~\eqref{second_IR1_solution} and taking the expectation value, we obtain\footnote{{Under the assumed initial conditions, the two independent integration momenta $k$ and $q$ are subject to the lower cutoff $\sigma a(N_\mathrm{s.h.})H$, whereas no separate lower cutoff is imposed on the composite momentum $|\mathbf k+\mathbf q|$.}}
\begin{align}
    &\langle\phi_\uir^{(2)\mathrm a}(N,\mathbf x)\rangle
    =\Braket{
    \varphi_\mathrm{hom}[\phi^{(2)\mathrm a}_\uir(N_\mathrm{s.h.});N]}\notag\\
    &\quad-\frac{\bar\pi(N)}2
    \int_{N_\mathrm{s.h.}}^N\dd N_1F(N_1)
    \int_{N_\mathrm{s.h.}}^{N_1}\dd N_2
    \frac{V_{\phi\phi\phi}(\bar\phi(N_2))}{F(N_2)\bar\pi(N_2)H^2}
    \times\int_{\sigma a(N_\mathrm{s.h.})H}^{\sigma a(N_2)H}
    \frac{\dd q}{q}\frac{q^3}{2\pi^2}P_q(N_2,N_2).
    \label{second_IR1_solution_mean}
\end{align}

Similarly, the mixed expectation value becomes
\begin{align}
    &\langle(\phi_\uir^{(1)}(N,\mathbf x))^2
    \phi_\uir^{(2)\mathrm a}(N,\mathbf x)\rangle
    \notag\\
    &=\Braket{\phi_\uir^{(1)}(N)^2\varphi_\mathrm{hom}[ \phi_{\mathrm{IR}}^{(2)\mathrm{a}}({N_\mathrm{s.h.}});N]}  + \langle(\phi^{(1)}_\uir(N))^2\rangle\com{\langle\phi_\uir^{(2)\mathrm{a}}(N)\rangle - \Braket{\varphi_\mathrm{hom}[\phi_{\mathrm{IR}}^{(2)\mathrm{a}}({N_\mathrm{s.h.}});N]}}\notag\\
    &-\bar\pi(N)
    \int_{N_\mathrm{s.h.}}^N\dd N_1F(N_1)
    \int_{N_\mathrm{s.h.}}^{N_1}\dd N_2
    \frac{V_{\phi\phi\phi}(\bar\phi(N_2))}{F(N_2)\bar\pi(N_2)H^2}\notag\\
    &\quad\times\int_{\mathbf k}\int_{\mathbf q}
    W\prn{\frac{k}{k_\sigma(N)}}
    W\prn{\frac{q}{k_\sigma(N)}}
    W\prn{\frac{|\mathbf k+\mathbf q|}{k_\sigma(N)}}
    W\prn{\frac{k}{k_\sigma(N_2)}}
    W\prn{\frac{q}{k_\sigma(N_2)}}
    P_k(N_2,N)P_q(N_2,N).
    \label{second_IR1_solution_3-pointpartial_final}
\end{align}
Here, we have used $W^2=W$ for the sharp window and the expectation value of the quadratic IR source,
\begin{align}
    &\langle\phi_\uir^{(1)}(N_2,\mathbf q)
    \phi_\uir^{(1)}(N_2,\mathbf k-\mathbf q)\rangle=(2\pi)^3\delta^{(3)}(\mathbf k)
    W\prn{\frac{q}{k_\sigma(N_2)}}^2P_q(N_2,N_2),
\end{align}

Furthermore, for Gaussian first-order perturbations, Wick's theorem gives
\begin{align}
    &\langle\phi^{(1)}(N,\mathbf k)\phi^{(1)}(N,\mathbf q)
    \phi^{(1)}(N_2,\mathbf p)\phi^{(1)}(N_2,\mathbf r)\rangle
    =(2\pi)^6\delta^{(3)}(\mathbf k+\mathbf q)
    \delta^{(3)}(\mathbf p+\mathbf r)
    P_k(N,N)P_p(N_2,N_2)\notag\\
    &\hspace{20mm}+(2\pi)^6\Big[
    \delta^{(3)}(\mathbf k+\mathbf p)\delta^{(3)}(\mathbf q+\mathbf r)
    +\delta^{(3)}(\mathbf k+\mathbf r)\delta^{(3)}(\mathbf q+\mathbf p)
    \Big]P_k(N_2,N)P_q(N_2,N).
    \label{Wick_expansion}
\end{align}
The first Wick contraction reproduces the disconnected contribution from the sourced part of $\phi_\uir^{(2)\mathrm a}$, which cancels against the corresponding term in Eq.~\eqref{3-point_definition}. The remaining two contractions yield the last term in Eq.~\eqref{second_IR1_solution_3-pointpartial_final}.

Combining the connected contributions and including the overall factor of three in Eq.~\eqref{3-point_definition}, we obtain
\begin{align}
    &\langle(\phi_\uir(N,\mathbf x)
    -\langle\phi_\uir(N,\mathbf x)\rangle)^3\rangle\big|_\mathrm a\notag\\
    &= 3\Braket{\phi_\uir^{(1)}(N)^2\varphi_\mathrm{hom}[ \phi_{\mathrm{IR}}^{(2)\mathrm{a}}({N_\mathrm{s.h.}});N]} - 3\Braket{\phi_\uir^{(1)}(N)^2}\Braket{\varphi_\mathrm{hom}[ \phi_{\mathrm{IR}}^{(2)\mathrm{a}}({N_\mathrm{s.h.}});N]}\notag\\
    &\quad-3\bar\pi(N)
    \int_{N_\mathrm{s.h.}}^N\dd N_1F(N_1)
    \int_{N_\mathrm{s.h.}}^{N_1}\dd N_2
    \frac{V_{\phi\phi\phi}(\bar\phi(N_2))}{F(N_2)\bar\pi(N_2)H^2}\notag\\
    &\quad\times\int_{\mathbf k}\int_{\mathbf q}
    W\prn{\frac{k}{k_\sigma(N)}}
    W\prn{\frac{q}{k_\sigma(N)}}
    W\prn{\frac{|\mathbf k+\mathbf q|}{k_\sigma(N)}}W\prn{\frac{k}{k_\sigma(N_2)}}
    W\prn{\frac{q}{k_\sigma(N_2)}}
    P_k(N_2,N)P_q(N_2,N).
    \label{second_IR1_solution_3-point_final}
\end{align}
The first two terms represent the connected contribution from the homogeneous solution, while the last term arises from nonlinear evolution after $N_\mathrm{s.h.}$.

\paragraph{Contribution labeled ``\(\mathrm{b}\)'': non-Markovian effects from IR--UV interactions} \mbox{}\\
We next consider Eq.~\eqref{second_IR_split2}, whose solution receives contributions from both the directly projected IR--UV interactions and the subsequent inflow of UV modes. Following the same procedure as for contribution ``a'' and applying Wick's theorem as in Eq.~\eqref{Wick_expansion}, we obtain the corresponding connected three-point contribution:
\begin{align}
    &\langle(\phi_\uir(N,\mathbf x)-\langle\phi_\uir(N,\mathbf x)\rangle)^3\rangle\big|_\mathrm b\notag\\
    &= 3\Braket{\phi_\uir^{(1)}(N)^2\varphi_\mathrm{hom}[ \phi_{\mathrm{IR}}^{(2)\mathrm{b}}({N_\mathrm{s.h.}});N]} - 3\Braket{\phi_\uir^{(1)}(N)^2}\Braket{\varphi_\mathrm{hom}[ \phi_{\mathrm{IR}}^{(2)\mathrm{b}}({N_\mathrm{s.h.}});N]}\notag\\
    &\quad-3\bar\pi(N)\int_{N_\mathrm{s.h.}}^N\dd N_1F(N_1)
    \int_{N_\mathrm{s.h.}}^{N_1}\dd N_2\frac{V_{\phi\phi\phi}(\bar\phi(N_2))}{F(N_2)\bar\pi(N_2)H^2}\notag\\
    &\quad\times\int_{\mathbf k}\int_{\mathbf q}
    W\prn{\frac{k}{k_\sigma(N)}}W\prn{\frac{q}{k_\sigma(N)}}
    W\prn{\frac{|\mathbf k+\mathbf q|}{k_\sigma(N)}}P_k(N_2,N)P_q(N_2,N)\notag\\
    &\quad\times\Bigg[W\prn{\frac{k}{k_\sigma(N_2)}}\com{1-W\prn{\frac{q}{k_\sigma(N_2)}}}
    +\com{1-W\prn{\frac{k}{k_\sigma(N_2)}}}W\prn{\frac{q}{k_\sigma(N_2)}}\Bigg].
    \label{second_IR2_solution_3-point_final}
\end{align}

\paragraph{Contribution labeled ``\(\mathrm{c}\)'': quadratic UV--UV interactions and non-white noise} \mbox{}\\
We also consider the UV--UV interactions, which generate both directly projected IR contributions and UV modes that subsequently enter the IR sector. The resulting second-order noise is quadratic in the Gaussian first-order UV modes and generally exhibits non-white correlations. Following the same procedure, we obtain
\begin{align}
    &\langle(\phi_\uir(N,\mathbf x)-\langle\phi_\uir(N,\mathbf x)\rangle)^3\rangle\big|_\mathrm c\notag\\
    &= 3\Braket{\phi_\uir^{(1)}(N)^2\varphi_\mathrm{hom}[ \phi_{\mathrm{IR}}^{(2)\mathrm{c}}({N_\mathrm{s.h.}});N]} - 3\Braket{\phi_\uir^{(1)}(N)^2}\Braket{\varphi_\mathrm{hom}[ \phi_{\mathrm{IR}}^{(2)\mathrm{c}}({N_\mathrm{s.h.}});N]}\notag\\
    &\quad-3\bar\pi(N)\int_{N_\mathrm{s.h.}}^N\dd N_1F(N_1)
    \int_{N_\mathrm{s.h.}}^{N_1}\dd N_2\frac{V_{\phi\phi\phi}(\bar\phi(N_2))}{F(N_2)\bar\pi(N_2)H^2}\int_{\mathbf k}\int_{\mathbf q}
    W\prn{\frac{k}{k_\sigma(N)}}W\prn{\frac{q}{k_\sigma(N)}}
    W\prn{\frac{|\mathbf k+\mathbf q|}{k_\sigma(N)}}\notag\\
    &\qquad\times\com{1-W\prn{\frac{k}{k_\sigma(N_2)}}}
    \com{1-W\prn{\frac{q}{k_\sigma(N_2)}}}
    P_k(N_2,N)P_q(N_2,N).
    \label{second_IR_NW_solution_3-point_final}
\end{align}

Finally, we compute the total three-point function of $\phi_{\mathrm{IR}}$ up to second order in perturbations. This is obtained by summing the three contributions, since $\phi_{\mathrm{IR}}^{(2)\mathrm{a}}$, $\phi_{\mathrm{IR}}^{(2)\mathrm{b}}$, and $\phi_{\mathrm{IR}}^{(2)\mathrm{c}}$ enter linearly in the three-point function at this perturbative order (see Eq.~\eqref{3-point_definition}). Although we have retained the general initial conditions so far, we now neglect the homogeneous contributions from the second-order UV modes by hand, restricting our analysis to nonlinear evolution generated on superhorizon scales after $N=N_\mathrm{s.h.}$. Specifically, {in Eqs.~\eqref{second_UV_solution a}-\eqref{second_UV_solution c}} we impose
\begin{align}
    \varphi_\mathrm{hom}[\phi_\uuv^{(2)i}(N_\mathrm{s.h.},k);N]=0\qquad k>\sigma a(N_\mathrm{s.h.}) H.\label{initial condition UV2}
\end{align}
We leave the incorporation of general second-order UV initial conditions, which may encode nonlinear effects generated during subhorizon evolution and around horizon crossing, to future work. Thus, combining the three contributions derived above, we obtain the following expression for the three-point correlator generated by nonlinear superhorizon evolution:
\begin{align} \nonumber
   &\langle \left(\phi_{\mathrm{IR}}(N)-\langle\phi_{\mathrm{IR}}(N)\rangle\right)^3\rangle\simeq  \mathrm{Eq.~}\eqref{second_IR1_solution_3-point_final} +\mathrm{Eq.~}\eqref{second_IR2_solution_3-point_final}+\mathrm{Eq.~}\eqref{second_IR_NW_solution_3-point_final} \\
   &=3\Braket{\phi_\uir^{(1)}(N)^2\varphi_\mathrm{hom}[ \phi_{\mathrm{IR}}^{(2)}(N_\mathrm{s.h.}(N));N]} - 3\Braket{\phi_\uir^{(1)}(N)^2}\Braket{\varphi_\mathrm{hom}[ \phi_{\mathrm{IR}}^{(2)}(N_\mathrm{s.h.}(N));N]}\notag\\
    &\quad    -3\bar\pi(N)\int_{N_\mathrm{s.h.}}^N\dd N_1F(N_1)
    \int_{N_\mathrm{s.h.}}^{N_1}\dd N_2\frac{V_{\phi\phi\phi}(\bar\phi(N_2))}{F(N_2)\bar\pi(N_2)H^2}\notag\\
    &\qquad\times\int_{\mathbf k}\int_{\mathbf q}
    W\prn{\frac{k}{k_\sigma(N)}}W\prn{\frac{q}{k_\sigma(N)}}
    W\prn{\frac{|\mathbf k+\mathbf q|}{k_\sigma(N)}}
    P_k(N_2,N)P_q(N_2,N).
   \label{second_IR_total_solution_3-point_final_before}
\end{align}
Note that, at each interaction time $N_2$, the input window functions sum to unity:
\begin{align}
    &W\prn{\frac{k}{k_\sigma(N_2)}}W\prn{\frac{q}{k_\sigma(N_2)}}
    +W\prn{\frac{k}{k_\sigma(N_2)}}\com{1-W\prn{\frac{q}{k_\sigma(N_2)}}}\notag\\
    &\quad+\com{1-W\prn{\frac{k}{k_\sigma(N_2)}}}W\prn{\frac{q}{k_\sigma(N_2)}}
    +\com{1-W\prn{\frac{k}{k_\sigma(N_2)}}}\com{1-W\prn{\frac{q}{k_\sigma(N_2)}}}=1.
    \label{input window completeness}
\end{align}
The window acting on the final output momentum $|\mathbf k+\mathbf q|$ is common to all three terms and does not cancel. For a monotonically increasing sharp cutoff, this can be understood from the identity
\begin{align}
    W\prn{\frac{|\mathbf k+\mathbf q|}{k_\sigma(N)}}
    &=W\prn{\frac{|\mathbf k+\mathbf q|}{k_\sigma(N_2)}}+\left[W\prn{\frac{|\mathbf k+\mathbf q|}{k_\sigma(N)}}
    -W\prn{\frac{|\mathbf k+\mathbf q|}{k_\sigma(N_2)}}\right].
    \label{output window history}
\end{align}
The first term accounts for modes already in the IR sector at the interaction time, while the second accounts for nonlinear modes that subsequently enter the IR sector through the second-order noise.

To isolate the nonlinear evolution generated after $N=N_\mathrm{s.h.}$, we further consider the case in which the homogeneous contributions associated with the second-order IR initial conditions vanish. We implement this restriction by assuming a Gaussian initial state and neglecting nonlinear interactions prior to $N_\mathrm{s.h.}$. As discussed in Appendix~\ref{sec:tree-initial-value}, this prescription can be implemented by setting the second-order perturbations of the canonical pair $\phi$ and $\Pi_\mathrm{c}$ to zero at $N=N_\mathrm{s.h.}$. For a spectator field on a fixed de Sitter background without derivative interactions, imposing $\varphi_\mathrm{hom}[\phi_\uir^{(2)i}(N_\mathrm{s.h.});N]=0$ is then consistent with the corresponding tree-level quantum calculation.\footnote{{In general, boundary conditions at the coarse-graining scale should be treated carefully, since the IR and UV sectors arise from a decomposition of the same underlying field. For a discussion within the path-integral approach, see, for example, Refs.~\cite{Tokuda:2017fdh,Tokuda:2018eqs}.}} However, the same prescription is not generally applicable to the inflaton, as we will discuss in Sec.~\ref{sec:nonlinear3}. Under these assumptions, the three-point correlator arising from nonlinear superhorizon evolution for the spectator field reduces to
\begin{align} \nonumber
   &\langle \left(\phi_{\mathrm{IR}}(N)-\langle\phi_{\mathrm{IR}}(N)\rangle\right)^3\rangle=-3\bar\pi(N)\int_{N_\mathrm{s.h.}}^N\dd N_1F(N_1)
    \int_{N_\mathrm{s.h.}}^{N_1}\dd N_2\frac{V_{\phi\phi\phi}(\bar\phi(N_2))}{F(N_2)\bar\pi(N_2)H^2}\notag\\
    &\qquad\times\int_{\mathbf k}\int_{\mathbf q}
    W\prn{\frac{k}{k_\sigma(N)}}W\prn{\frac{q}{k_\sigma(N)}}
    W\prn{\frac{|\mathbf k+\mathbf q|}{k_\sigma(N)}}
    P_k(N_2,N)P_q(N_2,N).
   \label{second_IR_total_solution_3-point_final}
\end{align}
Equation~\eqref{second_IR_total_solution_3-point_final} is one of the main results of this work. As we will demonstrate in the next section, it agrees with the result obtained from the in-in calculation in Eq.~\eqref{three point correlator inin}, within the same superhorizon approximation and under the same assumptions about the initial conditions. This agreement explicitly demonstrates the importance of including all three contributions: the contribution labeled ``$\mathrm{a}$", which would coincide with its counterpart in the conventional framework if the noise contributions arising from UV modes sourced by IR–IR interactions were neglected; the history-dependent IR–UV interactions, \ie~the non-Markovian contribution labeled ``$\mathrm{b}$"; and the quadratic UV–UV interactions generating non-white-noise contributions labeled ``$\mathrm{c}$". In particular, the latter two contributions, which are often neglected in the literature, enter at the same perturbative order as the contribution labeled ``$\mathrm{a}$", although they involve different ranges of scales. Their inclusion is therefore essential for reproducing the corresponding tree-level quantum result within the framework considered here.

\subsubsection{The inflaton including metric perturbations}\label{sec:nonlinear3}
Next, let us extend the formulation developed above for the nonlinear regime on a fixed de Sitter background to the case of the inflaton coupled to dynamical geometry. The classical dynamics of the inflaton at second order \cite{Cruces:2024pni} is given by 
\begin{align}
    &\dif{\phi^{(2)}}{N} = \pi^{(2)},\\
    &\dif{\pi^{(2)}}{N} + \prn{3 - \epsilon_1 + \frac{\epsilon_1\epsilon_2}{3 - \epsilon_1}} \pi^{(2)} + \prn{-\frac{3}{2}\epsilon_2 + \frac{1}{2} \epsilon_1\epsilon_2 - \frac{1}{4}\epsilon_2^2 - \frac{1}{2}\epsilon_2\epsilon_3 - \frac{\epsilon_1\epsilon_2^2}{2(3 - \epsilon_1)} } \phi^{(2)} \notag\\
    & - \frac{\epsilon_1}{2\bar\pi} \prn{ 4 - \frac{\epsilon_2}{3-\epsilon_1}}(\pi^{(1)})^2 + \frac{\epsilon_1}{\bar\pi}\prn{\epsilon_2 + \frac{\epsilon_2^2}{2} \frac{3+\epsilon_1}{(3 - \epsilon_1)^2}  + \frac{\epsilon_2\epsilon_3}{3 - \epsilon_1} } \pi^{(1)}\phi^{(1)} \notag\\
    & - \frac{\epsilon_2}{8\bar\pi}\prn{\epsilon_1\epsilon_2^2 \frac{9 + \epsilon_1}{(3 - \epsilon_1)^2} + 2\epsilon_2\epsilon_3\frac{3 + 2\epsilon_1}{3 - \epsilon_1} + 2\epsilon_3(3 - \epsilon_1) + 2\epsilon_3^2 + 2\epsilon_3\epsilon_4}(\phi^{(1)})^2=0,
\end{align}
or, equivalently,
\begin{align}
    &F\bar\pi\dif{}{N}\left[\frac{1}{F}\dif{}{N}\left(\frac{\phi^{(2)}}{\bar{\pi}}\right)\right]\notag\\
    &= \frac{\epsilon_1}{2\bar\pi} \prn{ 4 - \frac{\epsilon_2}{3-\epsilon_1}}(\pi^{(1)})^2 - \frac{\epsilon_1}{\bar\pi}\prn{\epsilon_2 + \frac{\epsilon_2^2}{2} \frac{3+\epsilon_1}{(3 - \epsilon_1)^2}  + \frac{\epsilon_2\epsilon_3}{3 - \epsilon_1} } \pi^{(1)}\phi^{(1)} \notag\\
    & \quad+ \frac{\epsilon_2}{8\bar\pi}\prn{\epsilon_1\epsilon_2^2 \frac{9 + \epsilon_1}{(3 - \epsilon_1)^2} + 2\epsilon_2\epsilon_3\frac{3 + 2\epsilon_1}{3 - \epsilon_1} + 2\epsilon_3(3 - \epsilon_1) + 2\epsilon_3^2 + 2\epsilon_3\epsilon_4}(\phi^{(1)})^2
\end{align}
Up to the validity of the separate universe approach which is employed in this paper, that is, leading order in the slow-roll (SR) parameters, the second order EoMs reduce to 
\begin{align}
    F\bar\pi\dif{}{N}\left[\frac{1}{F}\dif{}{N}\left(\frac{\phi^{(2)}}{\bar{\pi}}\right)\right]=\frac{3}{4\bar{\pi}}\epsilon_2\epsilon_3\left(\phi^{(1)}\right)^2. \label{eq:second-order-attractor}
\end{align}
The structure of this equation is the same as that in the fixed de Sitter case, Eq.~\eqref{Second order EoM deSitter}, except for the source terms on the right-hand side. Therefore, the IR modes can be analyzed following the same procedure as in the previous case, by decomposing the solution into three contributions labeled ``$\mathrm{a}$", ``$\mathrm{b}$", and ``$\mathrm{c}$". Thus, the correlator can be obtained from Eq.~\eqref{second_IR_total_solution_3-point_final_before}, derived for a fixed de Sitter background, by making the replacement $-\frac{1}{2}\frac{V_{\phi\phi\phi}}{H^2} \rightarrow \frac{3}{4\bar{\pi}}\epsilon_2\epsilon_3$, and we obtain
\begin{align}
   &\langle \left(\phi_{\mathrm{IR}}(N)-\langle\phi_{\mathrm{IR}}(N)\rangle\right)^3\rangle\notag\\
   &\simeq3\Braket{\phi_\uir^{(1)}(N)^2\varphi_\mathrm{hom}[ \phi_{\mathrm{IR}}^{(2)}(N_\mathrm{s.h.}(N));N]} - 3\Braket{\phi_\uir^{(1)}(N)^2}\Braket{\varphi_\mathrm{hom}[ \phi_{\mathrm{IR}}^{(2)}(N_\mathrm{s.h.}(N));N]}\notag\\
   &\quad +3\bar{\pi}(N)\int_{N_\mathrm{s.h.}}^N \dd N_1F(N_1)\int_{N_\mathrm{s.h.}}^{N_1}\dd N_2\frac{3\epsilon_2(N_2)\epsilon_3(N_2)}{2F(N_2)\bar{\pi}(N_2)^2}\notag\\
   &\qquad\times\int_{\mathbf{k}}\int_{\mathbf{q}}
       W\prn{\frac{k}{k_\sigma(N)}}W\prn{\frac{q}{k_\sigma(N)}}
       W\prn{\frac{|\mathbf{k}+\mathbf{q}|}{k_\sigma(N)}}\phi^{(1)}\left(k,N_2\right)\phi^{(1)}\left(k,N\right)\phi^{(1)}\left(q,N_2\right)\phi^{(1)}\left(q,N\right).
   \label{second_IR_total_solution_3-point_SR_before general}
\end{align}

If {we consider} attractor models at leading order in the SR parameters, Eq.~\eqref{second_IR_total_solution_3-point_SR_before general} can be computed {analytically}\footnote{For a sharp window $W(x)=\theta(1-x)$ and constant $H$, the integral in square brackets can be evaluated exactly. For $N-N_\mathrm{s.h.}\geq\ln 2$, it is given by
\begin{align}
&\int_{\sigma a(N_\mathrm{s.h.})H}^{\sigma a(N)H}\frac{\dd k}{k}\int_{\sigma a(N_\mathrm{s.h.})H}^{\sigma a(N)H}\frac{\dd q}{q}\frac12\int_{-1}^{1}\dd\mu\,W\left(\frac{\sqrt{k^2+q^2+2kq\mu}}{k_\sigma(N)}\right)\notag \\
&=(N-N_\mathrm{s.h.})^2-\frac{\pi^2}{12}+\frac14+\operatorname{Li}_2\left(e^{-(N-N_\mathrm{s.h.})}\right)-\frac12e^{-(N-N_\mathrm{s.h.})}.
\end{align} where $\operatorname{Li}_2(z)$ denotes the dilogarithm function, defined by
$\operatorname{Li}_2(z)=-\int_0^z \dd t\,\ln(1-t)/t$.} as
\begin{align}
   &\langle \left(\phi_{\mathrm{IR}}(N)-\langle\phi_{\mathrm{IR}}(N)\rangle\right)^3\rangle\notag\\
   &\simeq3\Braket{\phi_\uir^{(1)}(N)^2\varphi_\mathrm{hom}[ \phi_{\mathrm{IR}}^{(2)}(N_\mathrm{s.h.}(N));N]} - 3\Braket{\phi_\uir^{(1)}(N)^2}\Braket{\varphi_\mathrm{hom}[ \phi_{\mathrm{IR}}^{(2)}(N_\mathrm{s.h.}(N));N]}\notag\\
&\quad+\frac{3}{2\bar{\pi}_\mathrm{i}}
\left(\frac{\bar{H}_\mathrm{i}}{2\pi}\right)^4
\Bigg[
\int_{\sigma a(N_\mathrm{s.h.})H}^{\sigma a(N)H}
\frac{\dd k}{k}
\int_{\sigma a(N_\mathrm{s.h.})H}^{\sigma a(N)H}
\frac{\dd q}{q}
\frac12\int_{-1}^{1}\dd\mu\,
W\prn{
\frac{\sqrt{k^2+q^2+2kq\mu}}{k_\sigma(N)}
}
\Bigg]\notag\\
&\qquad\times
\com{
\epsilon_2(N)-\epsilon_2(N_\mathrm{s.h.})
-\frac{\epsilon_2'(N_\mathrm{s.h.})}
{F(N_\mathrm{s.h.})}
\int_{N_\mathrm{s.h.}}^N\dd N_1 F(N_1)
}. \label{second_IR_total_solution_3-point_SR_before}
\end{align}
where the last equality is obtained by expanding the expression to leading order in the SR parameters. In this calculation, we have used the solution for $\phi^{(1)}(N,k)$ in the SR approximation,
\begin{align}
    \phi^{(1)}(N,k) = {{k^{-3/2}\frac{\bar H_\mathrm{i}}{\sqrt{2}\bar\pi_\mathrm{i}}\com{1+\prn{\epsilon_1+\frac{\epsilon_2}{2}}\prn{\alpha - \log\prn{\frac{k}{\bar H_\mathrm{i}}}}-\epsilon_1}}} \times \bar\pi(N) \label{first UV attractor}
\end{align}
where $\alpha = 2 - \ln 2 - \gamma$ and $\gamma$ is the Euler–Mascheroni constant. We have also used the following relation, which holds at leading order in the SR parameters:
\begin{equation}
    \frac{3\epsilon_2(N_2)\epsilon_3(N_2)}{2F(N_2)}\simeq\dif{}{N_2}\left[\frac{1}{F(N_2)}\dif{}{N_2}\left(\frac{\epsilon_2(N_2)}{2}\right)\right].
\end{equation}

Let us now consider the initial conditions for the IR modes, following the same prescription as in the fixed de Sitter case. As discussed in the previous subsection, to implement our assumption that nonlinear effects generated prior to $N_\mathrm{s.h.}$ are neglected, we impose vanishing second-order perturbations of the canonical pair $\phi$ and $\Pi_\mathrm{c}$ at $N=N_\mathrm{s.h.}$. In the case of the inflaton, however, $\pi^{(2)}$ is not simply proportional to $\Pi_\mathrm{c}^{(2)}$, owing to the nonlinear contributions to the canonical momentum arising from derivative interactions and the associated boundary terms in Eq.~\eqref{cubic action flat IR}. Note that such contributions are absent for the spectator field on a fixed de Sitter background considered above. Consequently, as discussed in Appendix~\ref{sec:tree-initial-value}, imposing $\phi_\uir^{(2)}=\Pi_{\mathrm{c},\uir}^{(2)}=0$ at $N=N_\mathrm{s.h.}$ yields a nonvanishing homogeneous contribution:
\begin{align}
    &\varphi_\mathrm{hom}[\phi^{(2)}_\uir(N_\mathrm{s.h.});N] \notag\\
    &= -(\phi_\uir^{(1)}(N_\mathrm{s.h.}))^2\frac{\bar\pi(N)}{4 F(N_\mathrm{s.h.}) \bar\pi(N_\mathrm{s.h.})^2} \epsilon_1(N_\mathrm{s.h.})\com{2\epsilon_1(N_\mathrm{s.h.}) - 3\epsilon_2(N_\mathrm{s.h.}) } \int_{N_\mathrm{s.h.}}^N \dd N_1 F(N_1). \label{initial IR 2 SR}
\end{align}
Including this contribution, we obtain the following three-point correlator:

\begin{align}
    &\langle \left(\phi_{\mathrm{IR}}(N)
    -\langle\phi_{\mathrm{IR}}(N)\rangle\right)^3\rangle
    \simeq
    \frac{3}{2\bar{\pi}_\mathrm{i}}
    \left(\frac{\bar{H}_\mathrm{i}}{2\pi}\right)^4
    \int_{\sigma a(N_\mathrm{s.h.})H}^{\sigma a(N)H}
    \frac{\dd k}{k}
    \int_{\sigma a(N_\mathrm{s.h.})H}^{\sigma a(N)H}
    \frac{\dd q}{q}
    \frac12\int_{-1}^{1}\dd\mu\,
    W\prn{
    \frac{\sqrt{k^2+q^2+2kq\mu}}{k_\sigma(N)}
    }
    \notag\\
    &\quad\times
    \Bigg[
    \epsilon_2(N)-\epsilon_2(N_\mathrm{s.h.})-
    \frac{
    \epsilon_2(N_\mathrm{s.h.})\epsilon_3(N_\mathrm{s.h.})
    +\epsilon_1(N_\mathrm{s.h.})
    \com{2\epsilon_1(N_\mathrm{s.h.})
    -3\epsilon_2(N_\mathrm{s.h.})}
    }{F(N_\mathrm{s.h.})}
    \int_{N_\mathrm{s.h.}}^N\dd N_1F(N_1)
    \Bigg].
    \label{second_IR_total_solution_3-point_SR}
\end{align}
where we have used Eqs.~\eqref{first_2-point} and \eqref{first UV attractor}.

In the next section, we compare this result with the corresponding in-in calculation and clarify the predictions of stochastic inflation for the three-point correlator in attractor models at leading order in the SR expansion.

\subsection{Comments on the validity}

For clarity, let us summarize the approximations used above and discuss their validity in comparison with conventional statements in the literature for cosmological perturbations. 

For the IR modes, the literature works at leading order in the gradient expansion, so that spatial-derivative terms are neglected. In addition, the IR modes are assumed to behave classically, and they therefore consider the semiclassical limit in which the EoMs derived from the action principle hold. These assumptions are also adopted in our formulation. For the UV modes, the conventional formulation of the stochastic inflation in the literature retains only terms linear in perturbations and in the self-couplings {of the field}, so that their dynamics can be solved exactly within this approximation, leading to Gaussian statistics.

The key question is whether the IR modes can be quantitatively described beyond linear order in perturbations under the assumptions above. As we have shown in this paper, the answer is that the conventional formulation of stochastic inflation is justified only for linear perturbations. To correctly reproduce nonlinear cosmological perturbations, the following effects must be taken into account: non-Markovian effects and terms quadratic in the noise, both of which have been neglected in previous works, as discussed in \Sec{Sec:1}. The non-Markovian terms arise from the evolution of the UV modes on top of the IR background, namely from the coupling between the IR and UV modes, which is necessary for describing the system beyond linear order in perturbations. Moreover, as already implied by our previous work \cite{Cruces:2026yvs}, the terms quadratic in the noise are of the same perturbative order as the other contributions and therefore cannot be neglected. Thus, in order to compute the nonlinear dynamics quantitatively, we have proposed a reformulation of stochastic inflation that incorporates these two effects and have computed the resulting three-point correlators. In the next section, we will verify that the three-point correlator computed using the stochastic formalism in this section is indeed reproduced by the quantum in-in formalism.

We emphasize that the analytical results obtained in this section are valid only for nonlinear evolution on superhorizon scales. More specifically, our results apply when the gradient terms proportional to $k^2/(aH)^2$ in the UV equations \eqref{quantum_EOM_second} can be neglected and the superhorizon solutions \eqref{solution_MS_superhorizon} are applicable. We leave the inclusion of contributions from nonlinear interactions generated on subhorizon scales and around horizon crossing, as described by quantum field theory, to future work. These contributions are neglected by construction in the present analysis (see the discussion of initial conditions around Eq.~\eqref{initial condition UV2}).

{For completeness let us finally comment on the nonlinear extension from the perspective of resumming late-time IR divergences, which constitutes a well-founded motivation for this approach as discussed in Sec.~\ref{Sec:4}. Appendix~\ref{App:C} shows that IR–UV and UV–UV interactions remain relevant even in the late-time limit for superhorizon nonlinearity, modifying the coefficient of the leading IR divergence by a factor of three compared to the conventional formulation that includes only IR–IR interactions.}

\section{Comparison with quantum theoretical computations}\label{Sec:3}

Given a quantum system, the in-in formalism, or Schwinger--Keldysh formalism~\cite{Schwinger:1960qe,Keldysh:1964ud,Maldacena:2002vr}, allows one to compute correlation functions of arbitrary order directly from quantum field theory. Other approaches, such as the wave-function formalism~\cite{Maldacena:2011nz,Anninos:2014lwa} and the cosmological bootstrap~\cite{Arkani-Hamed:2018kmz}, can also be used to compute such quantities. In this section, we compare the results obtained from the first principles of quantum mechanics with those obtained from the stochastic formalism.

\subsection{Fixed de Sitter background}

Using the in-in formalism, the three-point correlator is given by
\begin{align}
    \langle\delta\phi(N,\textbf{x})^3\rangle&= \int _{\mathbf{k}_1,\mathbf{k}_2,\mathbf{k}_3}(2\pi)^3\delta(\mathbf{k}_1+\mathbf{k}_2+\mathbf{k}_3)B_\phi(k_1,k_2,k_3; N) \notag\\
    &= - i\int _{\mathbf{k}_1,\mathbf{k}_2,\mathbf{k}_3} e^{i(\mathbf{k}_1 + \mathbf{k}_2+ \mathbf{k}_3)\cdot \mathbf{x}} \int_{-\infty}^N \dd N' \Braket{\com{\delta\phi(N,\mathbf{k}_1)\delta\phi(N,\mathbf{k}_2)\delta\phi(N,\mathbf{k}_3), H_3(N')}}
\end{align}
with the cubic interaction Hamiltonian $H_3$. In the case of a fixed de Sitter background, it is given by
\begin{equation}
\hat{H}_3 = \frac{1}{6}\int \dd^3x \frac{a(N)^3}{H}V_{\phi\phi\phi}(\bar{\phi}(N))\hat{\delta\phi}^3.
    \label{interaction_hamiltonian_dS}
\end{equation}
Then, Wick's theorem yields
\begin{align}
      B_{\delta \phi}(\mathbf k_1,\mathbf k_2,\mathbf k_3;N)=2\operatorname{Im}\left[  u_{ k_1}(N)u_{k_2}(N)u_{k_3}(N) \int_{-\infty}^{N}\dd N_1\,  \frac{a^3(N_1)}{H}V_{\phi\phi\phi}(\bar\phi(N_1)) u_{ k_1}^*(N_1)u_{k_2}^*(N_1)u_{ k_3}^*(N_1)
  \right].
  \label{eq:summary-full-bispectrum}
\end{align}
where $u_{k_i}(N)$ denotes the mode function.

To compare this result with that obtained from the stochastic formalism in the previous section, let us consider the modes in the range $\sigma a(N_\mathrm{s.h.})H < k < \sigma a(N) H$ and decompose the bispectrum into two parts, $B_{\delta \phi}(N)=B_{\delta\phi,\rm sub/hc}(N)+B_{\delta\phi,\rm s.h.}(N)$:
\begin{align}
    B_{\delta\phi, \rm sub/hc}(N)&\equiv\int_{-\infty}^{N_\mathrm{s.h}} \dd N [\cdots],\quad B_{\delta\phi,\rm s.h.}(N)\equiv \int^{N}_{N_\mathrm{s.h}} \dd N [\cdots],  \label{def bispectrum iruv}
\end{align}
The first contribution, $B_{\delta\phi,\rm sub/hc}$, represents the intrinsic non-Gaussianity generated during subhorizon evolution and around horizon crossing, before the superhorizon approximation becomes valid. By contrast, $B_{\delta\phi,\rm s.h.}$ represents the contribution accumulated through nonlinear superhorizon evolution of modes in the range $\sigma a(N_\mathrm{s.h.})H<k<\sigma a(N)H$, after they have exited the horizon and entered the long-wavelength regime. This is precisely the contribution computed in the previous section in the regime where the $k$-dependence of the UV-mode evolution can be neglected, allowing us to use the superhorizon solutions for \Eq{first_UV_modified}
\begin{equation}
    u_k(N)=C_1(k)\bar{\pi}(N)+C_2(k)\bar{\pi}(N)\int^{N}F(N')\dd N',
    \label{solution_MS_superhorizon}
\end{equation}
and the bispectrum can then be written as follows (see the Appendix \ref{Sec:A}):
\begin{align}
    B_{\delta\phi,\mathrm{s.h.}}(k_1,k_2,k_3,N) =  -\int_{N_\mathrm{s.h.}}^N \dd N_1\frac{a(N_1)^3}{H} {V_{\phi\phi\phi}(\bar{\phi}(N_1))}G^{\mathrm R}_0(N,N_1)\com{P_{k_1}(N,N_1)P_{k_2} (N,N_1)+ (\mathrm{2\ perm})} \label{bispectrum superhorizon}
\end{align}
where
\begin{align}
    P_k(N,N_1) &\equiv \frac{1}{2}  \prn{G_k^>(N,N_1) + G_k^<(N,N_1)} = \operatorname{Re} \com{u_k(N)u_k^*(N_1)} \label{def P_k}\\
    G_k^{\mathrm R} (N,N_1) &\equiv -2\theta(N - N_1) \frac{1}{2i}\com{G^>_k(N,N_1) - G_k^<(N,N_1)} = - 2\theta(N - N_1) \operatorname{Im} \com{u_k(N)u_k^*(N_1)} \label{def G_k}\\
    &\xrightarrow{k\sim0\ \mathrm{with\ Eq.} \eqref{solution_MS_superhorizon}}  G_0^{\mathrm R}(N,N_1) =  \frac{\bar{\pi}(N)\bar{\pi}(N_1)}{2\mpl^2}\int_{N_1}^{N}F(N_2)\dd N_2 \label{G_0 value}
\end{align}
with
\begin{align}
    G_k^>(N,N_1) \equiv u_k(N)u_k^*(N_1),\quad G_k^<(N,N_1) \equiv u_k^*(N)u_k(N_1).
\end{align}
Note that the solution is valid only on sufficiently superhorizon scales, {In fact, note that the bispectrum obtained in \eqref{bispectrum superhorizon} is consistent with a local {type}, as it is expected from the fact that all modes are at superhorizon scales}. Accordingly, the $(G^{\mathrm R})^3$ term has been neglected in Eq.~\eqref{bispectrum superhorizon} because $G_k^{\mathrm R}$ is suppressed compared to $P_k$ in the $k\sim0$ limit.

In position space, using the bispectrum \eqref{bispectrum superhorizon}, the connected three-point correlator for IR modes with the window function can be expressed as
\begin{align}
    &\langle\delta\phi_\uir(N,\textbf{x})^3\rangle_\mathrm{con}=-3\bar{\pi}(N)\int_{N_\mathrm{s.h.}}^N \dd N_1 F(N_1)
      \int_{N_\mathrm{s.h.}}^{N_1}\dd N_2
      \frac{V_{\phi\phi\phi}(\bar{\phi}(N_2))}{F(N_2)\bar{\pi}(N_2)H^2}\notag\\
    &\quad\times\int_{\mathbf{k}}\int_{\mathbf{q}}W\prn{\frac{k}{k_\sigma(N)}}
      W\prn{\frac{q}{k_\sigma(N)}}
      W\prn{\frac{|\mathbf{k}+\mathbf{q}|}{k_\sigma(N)}}
      P_k(N_2,N)P_q(N_2,N).
      \label{three point correlator inin}
\end{align}
The remaining step is to specify the lower limit of the $k$-integration, which defines the IR cutoff. Since we focus on nonlinear superhorizon evolution beginning at $N=N_\mathrm{s.h.}$, we choose this cutoff to be $\sigma a(N_\mathrm{s.h.})H$. With the initial condition $\varphi_\mathrm{hom}[\phi_\uir^{(2)}(N_\mathrm{s.h.})]=0$, which eliminates nonlinear contributions generated prior to $N_\mathrm{s.h.}$ by construction, the above result agrees exactly with Eq.~\eqref{second_IR_total_solution_3-point_final}. This agreement demonstrates that both the non-Markovian and non-white-noise contributions must be incorporated into the stochastic formalism to consistently reproduce the quantum-theoretical result for nonlinear superhorizon evolution.

Let us emphasize once again what has been calculated here. Within the regime in which the $k$-dependent corrections can be neglected, all relevant UV modes are already close to or outside the horizon and can therefore be described by Eq.~\eqref{solution_MS_superhorizon}. In other words, the purely subhorizon dynamics contributing to the first term in Eq.~\eqref{def bispectrum iruv} is not captured by the present analysis. In fact, these contributions are neglected by imposing Eq.~\eqref{initial condition UV2} or IR ones as discussed above. Such dynamics can generate additional non-Gaussianity and constitutes a natural next step for future analysis.

\paragraph{Massless cubic interacting case.} As an explicit example, let us consider a massless scalar field with a cubic self-interaction $V(\phi)=\lambda\phi^3/3!$, for which $V_{\phi\phi\phi}=\lambda$ is constant. The free mode function is then given by
\begin{align}
    u_k(\tau) = \frac{H}{\sqrt{2k^3}} (1 + i k\tau) e^{- i k \tau}. \label{massless mode function}
\end{align}
The full bispectrum can be decomposed into two contributions by splitting the in-in time integral at $\tau_{\mathrm h}$. The first contribution, which includes nonlinear effects generated during subhorizon evolution and around horizon crossing, is given by
\begin{align}
    B_{\delta\phi,\rm sub/hc}(\tau)& = \frac{\lambda H^2}{4\prod_i k_i^3} \operatorname{Im} \com{\prod_{j=1}^3(1 + ik_j\tau)e^{-ik_j\tau}\int^{\tau_{\mathrm h}}_{-\infty } \dd \tau_1 \frac{e^{i (k_1+k_2+k_3) \tau_1}}{\tau_1^4} \prod_{i=1}^3 (1 - i k_i \tau_1)}\notag\\
    &\sim \frac{\lambda H^2}{4\prod_i k_i^3}
  \left[
  \frac13\sum_i k_i^3\{\gamma+\ln(-k_t\tau_{\mathrm h})\}
  +k_t\sigma_2-\frac49 k_t^3
  \right],
  \label{eq:benchmark-massless}
\end{align}
where $\tau_{\mathrm h}$ is the conformal time corresponding to $N_{\mathrm h}$, $k_t\equiv k_1+k_2+k_3$, and $\sigma_2\equiv k_1k_2 + k_2k_3 + k_3k_1$. Here, the last expression is obtained in the limit $|k_t\tau_{\mathrm h}|\ll1$. This contribution lies outside the scope of the present stochastic calculation and is therefore left for future work.

The second contribution arises from nonlinear superhorizon evolution and is given by
\begin{align}
    B_{\delta\phi,\rm s.h.}(\tau) &= \frac{\lambda H^2}{4\prod_i k_i^3} \operatorname{Im} \com{\prod_{j=1}^3(1 + ik_j\tau)e^{-ik_j\tau}\int_{\tau_{\mathrm h}}^{\tau } \dd \tau_1 \frac{e^{i (k_1+k_2+k_3) \tau_1}}{\tau_1^4} \prod_{i=1}^3 (1 - i k_i \tau_1)}\notag\\
    &\sim  \frac{\lambda H^2}{4\prod_i k_i^3}  \frac13\sum_i k_i^3 \com{\ln \left|\frac{\tau}{\tau_{\mathrm h}}\right|  + \frac{1}{3}\prn{1 - \frac{\tau^3}{\tau_\mathrm{h}^3}}}.
\end{align}
This is precisely the contribution investigated in the present work. Indeed, {this result can also be obtained from \eqref{3-point_cubic}}. Therefore, our reformulation of the stochastic formalism for nonlinear perturbations reproduces the local-type bispectrum generated by nonlinear superhorizon evolution for arbitrary momentum configurations, provided that all modes of interest are sufficiently far outside the horizon.

\subsection{Slow-roll inflation including metric perturbations}
We now turn to the inflaton in attractor models at leading order in the SR parameters. Since the result obtained in the previous section accounts only for nonlinear superhorizon evolution, we compute the corresponding inflaton bispectrum in the flat gauge at leading order in the gradient expansion, consistently with the approximation adopted throughout this work. For attractor models, the cubic action~\cite{Ballesteros:2024zdp} in the flat gauge at this order is given by 
\begin{align}
    S_{3, \mathrm{flat,IR}} &= \int \dd N\dd^3 x \dif{}{N}\com{\frac{\mpl^2 a^3 \epsilon_1 H \epsilon_2\epsilon_3}{6}}  \prn{\frac{\delta\phi}{\bar\pi}}^3  + S_\mathrm{b},\label{cubic action flat IR}
\end{align}
where the boundary term $S_\mathrm{b}$ is defined as
\begin{align}
    S_\mathrm{b} \equiv \int \dd N\dd^3 x \dif{}{N}\com{\frac{\mpl^2 a^3 \epsilon_1 H}{6} \epsilon_1(2 \epsilon_1 - 3\epsilon_2)  \prn{\frac{\delta\phi}{\bar\pi}}^3 }.
\end{align}
Using this action, the in-in formalism yields the following bispectrum at leading order in the SR parameters (see Appendix~\ref{Sec:A}):
\begin{align}
    &B_{\delta\phi,\mathrm{s.h.}}(N) =\prn{P_{k_1}P_{k_2} + P_{k_2}P_{k_3} + P_{k_3}P_{k_1}} \left[\frac{\epsilon_2(N) - \epsilon_2(N_\mathrm{s.h})}{2\bar\pi_\mathrm{i}} - \frac{\mpl^2 a^3(N_\mathrm{s.h}) \epsilon_1(N_\mathrm{s.h}) H(N_\mathrm{s.h})}{2\bar\pi_\mathrm{i}} \right.\notag\\
    &\qquad\left.\times\left\{\epsilon_2(N_\mathrm{s.h})\epsilon_3(N_\mathrm{s.h}) + \epsilon_1(N_\mathrm{s.h}) \prn{2 \epsilon_1(N_\mathrm{s.h}) - 3\epsilon_2(N_\mathrm{s.h})}\right\} \int_{N_\mathrm{s.h}}^N \frac{\dd N_2}{\mpl^2 a^3 \epsilon_1 H}\right].\label{second_IR_total_solution_3-point_SR_inin}
\end{align}
Integrating this bispectrum over the same superhorizon momentum range, while retaining all three final-time window functions, yields the same three-window momentum structure as Eq.~\eqref{three point correlator inin}, rather than a factorized square of momentum integrals. The nonlinear-evolution contribution is consistent with the window-resolved stochastic expression in Eq.~\eqref{second_IR_total_solution_3-point_SR_before}. Unlike in the spectator case, the boundary contribution also depends on the nonlinear canonical-momentum initial data. Agreement of the full correlator with Eq.~\eqref{second_IR_total_solution_3-point_SR} requires both the condition \eqref{initial IR 2 SR} and consistent treatment of second-order UV initial data that subsequently enter the final IR band.

\section{Summary and discussion}\label{Sec:4}

\subsection{Assessment and lessons learned from the nonlinear extension}
We have formulated a nonlinear extension of the stochastic inflation based on a heuristic Langevin description, both for a spectator field on a fixed de Sitter background and for a slow-rolling inflaton. In this formulation, we consistently incorporate contributions that are usually omitted in the literature, namely non-Markovian effects and terms quadratic in the noise. As shown in \Sec{Sec:3}, including these contributions leads to three-point correlators that agree with those obtained from quantum theoretical computations such as the in-in formalism. This comparison implies that, in our reformulation of the stochastic formalism, the bispectrum {obtained} only from the contributions of the superhorizon nonlinear evolution (neglecting the contribution from subhorizon or horizon crossing) is the local-type contribution for arbitrary configurations of $\mathbf{k}_i$, provided that the modes of interest, whose wavenumbers are below the coarse-graining cutoff, are taken into account (see \Eqs{second_IR_total_solution_3-point_final},\eqref{bispectrum superhorizon},\eqref{three point correlator inin} or \Eqs{second_IR_total_solution_3-point_SR},\eqref{second_IR_total_solution_3-point_SR_inin}). Although our analysis is restricted to superhorizon evolution, with subhorizon dynamics neglected by construction, our results demonstrate that both the quadratic-noise terms and the non-Markovian terms must be included in order to quantitatively describe nonlinear perturbations.

Note that, in a genuinely free theory, the nonlinear modifications to the conventional stochastic formalism discussed here are unnecessary, since there are no interactions that generate either non-Markovian contributions or quadratic-noise corrections. This simply reflects the fact that free theories are straightforward to solve within quantum field theory. Indeed, in a free quantum field theory, a Fock space can be constructed explicitly, and, for a Gaussian vacuum state, all field correlation functions are completely determined by the two-point function through Wick's theorem. In this sense, there is no compelling theoretical necessity to introduce the stochastic formalism since the relevant observables can instead be computed directly within quantum field theory.\footnote{In the free theory, a nonzero mass regulates the IR divergences associated with secular growth without the need to introduce the stochastic formalism. On the other hand, a massless, self-interacting spectator field on a fixed de Sitter background exhibits secularly growing contributions in quantum-field-theoretic perturbation theory, providing a genuine motivation for the stochastic formalism, as discussed in next subsection.} 

Let us also comment on inflationary models with sufficiently flat potentials, including those exhibiting ultra-slow-roll (USR) phases or inflection points. In these cases, the inflaton potential may become approximately or even exactly constant, which might at first sight suggest that the theory becomes free. However, this is not the case once the coupling between the inflaton and dynamical gravity is taken into account. Even for an exactly flat potential, the action for perturbations in the spatially flat gauge contains interaction terms beyond quadratic order due to gravitational couplings. Our explicit analysis is restricted to attractor models at leading order in the SR parameters and is performed within the separate-universe approximation, which neglects nonlocal contributions and may therefore deviate from the genuine long-wavelength limit of the linear perturbation equations~\cite{Rigopoulos:2003ak,Cruces:2021iwq,Cruces:2022imf,Tanaka:2021dww,Tanaka:2024mzw}. Nevertheless, the need to incorporate non-Markovian contributions and terms quadratic in the noise may also extend to USR-like models. Consequently, our analysis {implies} that the conventional stochastic description of USR-like models may be justified only at linear order and cannot, in general, be assumed to provide a quantitative description of nonlinear perturbations.

There are, however, several subtleties associated with our formulation. As mentioned in the main text, the contribution from the subhorizon or the horizon crossing such as the first term in \Eq{def bispectrum iruv} should be analyzed in our formulation and thus we should clarify which kind of the bispetrum can be fully computed by the stochastic inflation. Moreover, as emphasized above, the heuristic derivation based on Langevin equations has not been rigorously established from first principles. It is therefore important to investigate how the quadratic-noise and non-Markovian terms appearing in our formulation can be derived within quantum field theory. Such quantum-field-theoretical derivations have been well developed for fields on de Sitter backgrounds using, for example, the density matrix, the wave function, or dynamical renormalization-group techniques, although their primary purpose has typically been to resum the IR secular growth that dominates at late times. Establishing the precise connection between these approaches and the formulation developed here is left for future work.

In addition, in the case of the inflaton, we might consider the following noise that could contribute the final results:
\begin{align} \nonumber 
 \Xi^{(2)\mathrm{d}}_{\phi}(N)=&
 \frac{\partial}{\partial k}\left(\sqrt{\frac{k^3}{2\pi^2}\abs{ \phi_\mathrm{UV}^{(1)}(N,k)}^2}\right)_{k=\bar k_\sigma(N)}\left(\frac{\partial \bar k_\sigma(N)}{\partial \bar{\phi}} \phi^{(1)}_{\mathrm{IR}}+\frac{\partial \bar k_\sigma(N)}{\partial \bar{\pi}} \pi^{(1)}_{\mathrm{IR}}\right)\xi(N)\,,\\
 \Xi^{(2)\mathrm{d}}_{\pi}(N)=&
 \frac{\partial}{\partial k}\left(\sqrt{\frac{k^3}{2\pi^2}\abs{ \pi_\mathrm{UV}^{(1)}(N,k)}^2}\right)_{k=\bar k_\sigma(N)}\left(\frac{\partial \bar k_\sigma(N)}{\partial \bar{\phi}} \phi^{(1)}_{\mathrm{IR}}+\frac{\partial \bar k_\sigma(N)}{\partial \bar{\pi}} \pi^{(1)}_{\mathrm{IR}}\right)\xi(N)\,,
 \label{noises_Markovian}
\end{align}
which have the multiplicative and Markovian nature of the noise, namely $\Xi^{(2)\mathrm{d}}_{i}(N)=\left(f_i(N)\phi_{\mathrm{IR}}^{(1)}+g_i(N)\pi_{\mathrm{IR}}^{(1)}\right)\xi(N)$. These might be able to be understood as the modulation of the cutoff $k=\sigma a H$ for higher order perturbations. If, in the cutoff, the Hubble parameter $H(\bar\phi,\bar\pi)$ through dynamical gravity is taken to be the local Hubble parameter $H_\uir(\phi_\uir,\pi_\uir)$, its perturbative modulation can generate an additional contribution through the $k$-dependence of the noise, of the same perturbative order as the other second-order contributions. However, the superhorizon evolution of the inflaton three-point correlator is correctly reproduced by our formulation without including such cutoff-induced contributions, as shown in Eqs.~\eqref{second_IR_total_solution_3-point_SR} and \eqref{second_IR_total_solution_3-point_SR_inin}. This might suggest that the Hubble parameter entering the cutoff may be more appropriately chosen as the unperturbed background quantity. A more precise understanding of this point is also left for future work.

\subsection{Reconsidering the role of and motivation for the stochastic approach}

For clarity, let us comment on the role of the stochastic approach in several perspectives. Historically, the stochastic formalism was introduced to provide an intuitive classical picture in which the dynamics can be computed using classical methods, although both the governing equations and the classicality of the description itself must be justified by the underlying quantum theory (see footnotes~\ref{footnote:decoherence},\ref{footnote:CCB}). 

More importantly, from a theoretical perspective, the stochastic formalism provides a means of resumming infrared IR secular growth, particularly for spectator fields with non-derivative interactions. In such cases, when secular effects render conventional perturbative quantum-field-theoretic calculations unreliable, a nonperturbative treatment becomes necessary, for which the stochastic formalism for a spectator field provides a well-motivated approach. On the other hand, in the case of the inflaton including gravitational backreaction, the situation is different. Since all interactions for the curvature perturbations are derivative couplings, there is no such secular divergence in principle. Besides, {due to diffeomorphism invariance,} the constant mode of the curvature perturbation that could generate IR divergence can be absorbed into the locally measured background~\cite{Urakawa:2010it,Urakawa:2010kr,Tanaka:2011aj,Pajer:2013ana}. Therefore, for the inflaton, there is no potential reason to introduce the stochastic formalism, which is the decent description for secular growth IR divergences. 

Nevertheless, independently of this well-founded theoretical motivation, one may attempt to construct a stochastic formalism capable of describing nonlinear cosmological perturbations. As demonstrated in this paper, quantitatively describing nonlinear cosmological perturbations requires going beyond the conventional stochastic treatment by incorporating non-Markovian effects and terms quadratic in the noise. Within our formulation, these contributions can be systematically incorporated only order by order in perturbation theory, casting doubt on claims in the literature that stochastic inflation provides a nonperturbative description of cosmological perturbations. As discussed in Sec.~\ref{Sec:2linear}, the conventional predictions for nonlinear perturbations are merely extrapolations of the linear description and are quantitatively incorrect due to contributions neglected at the same perturbative order. Nevertheless, one cannot entirely rule out the possibility that some qualitative features, such as the proposed exponential tails,\footnote{Indeed, the $\delta N$ formalism~\cite{Starobinsky:1982ee,Starobinsky:1986fxa,Sasaki1996,Sasaki:1998ug,Lyth:2004gb} can predict exponential tails in non-attractor models independently of those suggested by the conventional stochastic approach, although the coefficients in the exponents differ~\cite{Pi:2022ysn,Cai:2018dkf,Biagetti:2018pjj,Biagetti:2021eep}.} remain valid. However, such features remain mere speculation unless their physical validity is established within a consistent nonlinear formulation that incorporates the missing contributions. {The quantitative accuracy of the tail is not established by the leading stochastic truncation alone.}

In conclusion, our analysis invites us to reconsider the role and necessity of stochastic inflation in describing nonlinear cosmological perturbations, particularly those of the inflaton. In the setting considered here, the original motivation based on the resummation of IR secular divergences is absent for inflaton perturbations, owing to the structure of their derivative and gravitational interactions, as discussed above.\footnote{Moreover, several ingredients of the stochastic formulation for the inflaton remain to be rigorously established, including the treatment of the constraint equations in both the IR and UV sectors and the appropriate choice of the coarse-graining scale.} Nevertheless, the stochastic approach might offer the advantage of making field dynamics easier to understand through a simple and intuitive classical picture, albeit at a heuristic level, in line with its historical motivation. However, as our analysis illustrates, this intuitive understanding does not translate into a clear advantage in quantitative predictive power over first-principles quantum-field-theoretic methods, such as the in-in formalism. Even if its qualitative extrapolations eventually acquire physical meaning through justification by the underlying quantum theory, the stochastic formulation would remain a complicated computational alternative that obscures aspects of the underlying quantum dynamics.

\section*{Acknowledgments}
This work is supported by the National Key Research and Development Program of China
Grant No. 2021YFC2203004.  D.C. is supported by the
78th Batch of General Grants of the China Postdoctoral Science Foundation No. 2025M783447. T.K. acknowledges financial support from the Institute for Basic Science (IBS) under the project code, IBS-R018-D3 and JST SPRING, Japan Grant Number JPMJSP2180. The authors would also like to thank Guillermo Ballesteros, Cristiano Germani, Swagat Saurav Mishra, Gonzalo Palma, Gerasimos Rigopoulos, Teruaki Suyama, Spyros Sypsas, Fumio Uchida, Yusuke Yamada, Masahide Yamaguchi, Yuhang Zhu for useful discussions.

\appendix

\section{In-in computation for the three-point correlator}\label{Sec:A}

In this appendix, we present the detailed in-in calculations of the three-point correlators used in \Sec{Sec:3}, considering both a spectator field on a fixed de Sitter background and a slow-roll inflaton including metric perturbations.

\subsection{De Sitter background}

We begin by decomposing the unequal-time product of mode functions into its real and imaginary parts:
\begin{equation}
u_{k_1}(N)u_{k_1}^{*}(N_1)=\operatorname{Re}\left(u_{k_1}(N)u_{k_1}^{*}(N_1)\right)+i\operatorname{Im}\left(u_{k_1}(N)u_{k_1}^{*}(N_1)\right).
\end{equation}
Consequently, for $N>N_1$, we obtain
\begin{align} \nonumber
&\operatorname{Im}\left(u_{k_1}(N)u_{k_1}^{*}(N_1)u_{k_2}(N)u_{k_2}^{*}(N_1)u_{k_3}(N)u_{k_3}^{*}(N_1)\right) \nonumber\\
&=-\frac{1}{2}\prn{G^\mathrm{R}_{k_1}P_{k_2}P_{k_3}+P_{k_1}G^\mathrm{R}_{k_2}P_{k_3}+P_{k_1}P_{k_2}G^\mathrm{R}_{k_3}} + \frac{1}{8}G^\mathrm{R}_{k_1}G^\mathrm{R}_{k_2}G^\mathrm{R}_{k_3}
\label{decomposition_Im_part}
\end{align}
where $P_k$ and $G_k^\mathrm{R}$ are defined in Eqs.~\eqref{def P_k} and \eqref{def G_k}, respectively. The retarded Green's function $G_k^{\rm R}$ satisfies
\begin{equation}
\left[
\partial_N^2+3\partial_N+\frac{k^2}{a^2H^2}+\frac{V_{\phi\phi}(N)}{H^2}
\right]G_k^{\rm R}(N,N_1)
=\frac{\delta(N-N_1)}{a^3(N_1)H},
\label{eq:GR-eom}
\end{equation}
where the normalization follows from the Wronskian condition \eqref{eq:Wronskian}. Note that no long-wavelength approximation has been employed so far.

We next evaluate $\operatorname{Im}\left(u_k(N)u_k^{*}(N_1)\right)$ using the superhorizon solution for the mode function given in Eq.~\eqref{solution_MS_superhorizon}. This yields
\begin{equation}
\operatorname{Im}\left(u_k(N)u_k^{*}(N_1)\right)=\operatorname{Im}\left(C_2(k)C_1^{*}(k)\right)\bar{\pi}(N)\bar{\pi}(N_1)\int_{N_1}^{N}F(N_2)\dd N_2.
\label{Imaginary}
\end{equation}
The coefficient $\operatorname{Im}\left(C_2(k)C_1^{*}(k)\right)$ is determined by the Wronskian normalization condition for the quantum mode functions:
\begin{equation}
u_k(N)\left(\frac{d u_k(N)}{dN}\right)^{*}-\left(\frac{d u_k(N)}{dN}\right)u_k(N)^{*}=\frac{i}{a(N)^3H}.
\label{eq:Wronskian}
\end{equation}
Substituting the superhorizon solution \eqref{solution_MS_superhorizon} into this condition, we find
\begin{equation}
-2i\operatorname{Im}(C_2(k)C_1^{*}(k))F(N)\bar{\pi}(N)^2=\frac{i}{a(N)^3H}
\qquad \longrightarrow \qquad
\operatorname{Im}(C_2(k)C_1^{*}(k))=-\frac{1}{4\mpl^2},
\end{equation}
where we have used $F\equiv \frac{2\mpl^2}{a^3 \bar{\pi}^2 H}$. Consequently, Eq.~\eqref{Imaginary} reduces to
\begin{equation}
\operatorname{Im}\left(u_k(N)u_k^{*}(N_1)\right)=-\frac{\bar{\pi}(N)\bar{\pi}(N_1)}{4\mpl^2}\int_{N_1}^{N}F(N_2)\dd N_2,
\label{def of I0}
\end{equation}
which is directly related to the retarded Green's function in Eq.~\eqref{G_0 value}.

In the superhorizon regime, the last term in Eq.~\eqref{decomposition_Im_part}, which contains three retarded Green's functions, is subleading and can therefore be neglected. Furthermore, since the retarded Green's function becomes independent of $k$ at leading order in the gradient expansion, the momentum-conserving Dirac delta function allows us to perform the momentum integration associated with each retarded Green's function. The three remaining contributions are identical after relabeling the integration variables. We thus obtain
\begin{align}
\langle\delta\phi_\mathrm{IR}(\mathbf{x},N)^3\rangle_\mathrm{con}&= -\frac{3\bar{\pi}(N)}{2\mpl^2}
\int\frac{\dd^3 k_1 \dd^3 k_2}{(2\pi)^6}
W\prn{\frac{k_1}{k_\sigma(N)}}
W\prn{\frac{k_2}{k_\sigma(N)}}
W\prn{\frac{|\mathbf{k}_1+\mathbf{k}_2|}{k_\sigma(N)}}
\notag\\
&\quad\times
\int_{N_\mathrm{s.h.}}^N\dd N_1
\frac{a(N_1)^3}{H}
\bar{\pi}(N_1)V_{\phi\phi\phi}(\bar{\phi}(N_1))
\int_{N_1}^{N}\dd N_2
F(N_2)P_{k_1}(N,N_1)P_{k_2}(N,N_1).
\label{3-point-in-in_2}
\end{align}
Finally, using the identity
\begin{equation}
\frac{a(N)^3}{2\mpl^2H}\bar{\pi}(N)=\frac{1}{F(N)\bar{\pi}(N)H^2}
\end{equation}
and exchanging the order of integration according to
\begin{equation}
\int_{N_\mathrm{s.h.}}^{N}\dd N_1 f(N_1)\int_{N_1}^{N}\dd N_2g(N_2)=\int_{{N_\mathrm{s.h.}}}^{N}\dd N_1 g(N_1)\int_{{N_\mathrm{s.h.}}}^{N_1}\dd N_2f(N_2),
\end{equation}
we arrive at the expression given in Eq.~\eqref{three point correlator inin}:
\begin{align}
\langle\delta\phi_\mathrm{IR}(\mathbf{x},N)^3\rangle_\mathrm{con}&= -3\bar{\pi}(N)
\int_{N_\mathrm{s.h.}}^N \dd N_1 F(N_1)
\int_{N_\mathrm{s.h.}}^{N_1}\dd N_2
\frac{V_{\phi\phi\phi}(\bar{\phi}(N_2))}
{F(N_2)\bar{\pi}(N_2)H^2}
\notag\\
&\quad\times
\int\frac{\dd^3 k\,\dd^3 q}{(2\pi)^6}
W\prn{\frac{k}{k_\sigma(N)}}
W\prn{\frac{q}{k_\sigma(N)}}
W\prn{\frac{|\mathbf{k}+\mathbf{q}|}{k_\sigma(N)}}
P_k(N_2,N)P_q(N_2,N).
\end{align}

\subsection{Inflaton in the flat gauge}

The in-in computation of the bispectrum on superhorizon scales proceeds essentially in the same way as in the preceding subsection. At leading order in the gradient expansion, the linear mode function satisfies
\begin{align}
    \dif{}{N} \com{\frac{a^3\epsilon_1 H}{\bar\pi}\prn{\dif{u_k}{N} - \frac{\epsilon_2}{2}u_k}} = \dif{}{N} \com{{a^3\epsilon_1 H}\dif{}{N}\prn{\frac{u_k}{\bar\pi}}} =0,
\end{align}
where we have used $\bar\pi'/\bar\pi=\epsilon_2/2$. The mode function is subject to the Wronskian normalization condition
\begin{align}
    2 \mpl^2 a^3 \epsilon_1 H \com{\frac{u_k}{\bar\pi} \dif{}{N}\prn{\frac{u_k^*}{\bar\pi}} - \frac{u_k^*}{\bar\pi} \dif{}{N}\prn{\frac{u_k}{\bar\pi}}} = i.
\end{align}
On superhorizon scales, the growing mode of $u_k/\bar\pi$ is constant in attractor models, while the decaying mode becomes negligible in its amplitude. Nevertheless, the latter must be retained when evaluating the imaginary part of the unequal-time product, whose normalization is fixed by the Wronskian condition. We thus obtain
\begin{align}
    \operatorname{Im} \com{\frac{u_k(N)}{\bar\pi(N)} \frac{u_k^*(N_\mathrm{s.h.})}{\bar\pi(N_\mathrm{s.h.})}} = -\frac{1}{4} \int_{N_\mathrm{s.h.}}^N  \frac{\dd N_2}{\mpl^2a^3(N_2) \epsilon_1(N_2) H(N_2)},
\end{align}
which implies
\begin{align}
    \operatorname{Im} \com{\frac{u_k(N)}{\bar\pi(N)} \dif{}{N_\mathrm{s.h.}}\prn{\frac{u_k^*(N_\mathrm{s.h.})}{\bar\pi(N_\mathrm{s.h.})}} } =  \frac{1}{4\mpl^2a^3(N_\mathrm{s.h.}) \epsilon_1(N_\mathrm{s.h.}) H(N_\mathrm{s.h.})}.
\end{align}

Using the cubic action on superhorizon scales given in Eq.~\eqref{cubic action flat IR}, including its boundary contribution, the in-in formalism yields
\begin{align}
    &B_{\delta\phi,\mathrm{s.h.}}(N) =\prn{P_{k_1}P_{k_2} + P_{k_2}P_{k_3} + P_{k_3}P_{k_1}} \left[\frac{\epsilon_2(N) - \epsilon_2(N_\mathrm{s.h})}{2\bar\pi_\mathrm{i}} - \frac{\mpl^2 a^3(N_\mathrm{s.h}) \epsilon_1(N_\mathrm{s.h}) H(N_\mathrm{s.h})}{2\bar\pi_\mathrm{i}} \right.\notag\\
    &\qquad\left.\times\left\{\epsilon_2(N_\mathrm{s.h})\epsilon_3(N_\mathrm{s.h}) + \epsilon_1(N_\mathrm{s.h}) \prn{2 \epsilon_1(N_\mathrm{s.h}) - 3\epsilon_2(N_\mathrm{s.h})}\right\} \int_{N_\mathrm{s.h}}^N \frac{\dd N_2}{\mpl^2 a^3 \epsilon_1 H}\right], \label{B deltaphi IR}
\end{align}
which reproduces the result quoted in the main text at leading order in the slow-roll (SR) parameters. 

To examine the superhorizon evolution, we consider the difference between the bispectra evaluated at two superhorizon times, $N'$ and $N$, keeping the momenta and $N_\mathrm{s.h.}$ fixed. We obtain
\begin{align}
    &\Delta B_{\delta\phi,\mathrm{s.h.}}(N',N)\equiv B_{\delta\phi,\mathrm{s.h.}} (N') - B_{\delta\phi,\mathrm{s.h.}}(N) \notag\\
    &= \prn{P_{k_1}P_{k_2} + P_{k_2}P_{k_3} + P_{k_3}P_{k_1}} \left[\frac{\epsilon_2(N') - \epsilon_2(N)}{2\bar\pi_\mathrm{i}} - \frac{\mpl^2 a^3(N_\mathrm{s.h}) \epsilon_1(N_\mathrm{s.h}) H(N_\mathrm{s.h})}{2\bar\pi_\mathrm{i}} \right.\notag\\
    &\qquad\left.\times\left\{\epsilon_2(N_\mathrm{s.h})\epsilon_3(N_\mathrm{s.h}) + \epsilon_1(N_\mathrm{s.h}) \prn{2 \epsilon_1(N_\mathrm{s.h}) - 3\epsilon_2(N_\mathrm{s.h})}\right\} \prn{\int_{N}^\infty \frac{\dd N_2}{\mpl^2 a^3 \epsilon_1 H} - \int_{N'}^\infty \frac{\dd N_2}{\mpl^2 a^3 \epsilon_1 H}}\right]\notag\\
    &\simeq \prn{P_{k_1}P_{k_2} + P_{k_2}P_{k_3} + P_{k_3}P_{k_1}}\frac{\epsilon_2(N') - \epsilon_2(N)}{2\bar\pi_\mathrm{i}},
\end{align}
where the decaying-mode contribution has been neglected in the last line, as appropriate for the growing-mode regime of attractor models. On the other hand, the nonlinear gauge transformation
\begin{align}
    \zeta= -\frac{\delta\phi}{\bar\pi} + \frac{\epsilon_2}{4}\prn{-\frac{\delta\phi}{\bar\pi}}^2
\end{align}
relates the bispectra in the flat and comoving gauges. At the order considered here, this relation gives
\begin{align}
    B_\zeta(N)  = \frac{-1}{\bar\pi_\mathrm{i}^3} B_{\delta\phi} (N) + \frac{\epsilon_2(N)}{2\bar\pi_\mathrm{i}^4} \prn{P_{k_1}P_{k_2} + P_{k_2}P_{k_3} + P_{k_3}P_{k_1}}.
\end{align}
Consequently, the change in the curvature-perturbation bispectrum between the two superhorizon times is
\begin{align}
    \Delta B_{\zeta,\mathrm{s.h.}}(N',N)
    &= \frac{-1}{\bar\pi_\mathrm{i}^3} \Delta B_{\delta\phi,\mathrm{s.h.}} (N',N) + \frac{\epsilon_2(N')-\epsilon_2(N)}{2\bar\pi_\mathrm{i}^4} \prn{P_{k_1}P_{k_2} + P_{k_2}P_{k_3} + P_{k_3}P_{k_1}}
    \simeq 0.
\end{align}
Thus, the time dependence of the flat-gauge bispectrum cancels that arising from the nonlinear gauge transformation, up to the decaying-mode contribution. This is consistent with the conservation of the growing superhorizon curvature perturbation in attractor models. Indeed, as in Eq.~\eqref{eq:second-order-EOM-bispectrum}, this conservation can also be verified directly from the bulk equations of motion, or equivalently from the Heisenberg equations, using the superhorizon attractor solutions:
\begin{align}
    \dif{}{N} \com{a^3 \epsilon_1 H\dif{}{N}\prn{-\frac{\delta \phi}{\bar\pi} + \frac{\epsilon_2}{4}\prn{-\frac{\delta\phi}{\bar\pi}}^2}} =\dif{}{N} \com{a^3 \epsilon_1 H\dif{\zeta}{N}} =0.
\end{align}
This equation admits a constant growing-mode solution for $\zeta$, while the other independent solution is decaying in attractor models, consistently with the bispectrum calculation above.

\section{Tree-level evolution as an initial-value problem}
\label{sec:tree-initial-value}
\subsection{General formulation}
An alternative approach to computing quantum correlators is to formulate tree-level evolution in terms of classical equations of motion, with boundary conditions determined by the quantum state.~\cite{Weinberg:2008mc}. This formulation is particularly useful in our setup because it provides a direct bridge to stochastic inflation, where the classical equations of motion are the central objects. We therefore reformulate the in-in computation of the correlators in this language. The system is prepared in the Bunch-Davies vacuum in the asymptotic past and subsequently evolves according to an explicitly time-dependent action. We introduce an intermediate time $N_0$ and encode all information about the evolution prior to $N_0$ into the quantum state defined on the $N=N_0$ hypersurface. Our main interest is then the subsequent evolution from $N_0$ to a later time $N$. At tree level, the evolution of the wave functional can be evaluated through classical solutions satisfying the equations of motion and state-dependent boundary conditions at $N_0$. In this way, the usual in-in computation can be recast as a classical evolution problem with quantum-mechanical information entering through the initial boundary conditions.

In what follows, we employ the path-integral formulation, in which the saddle-point configurations satisfy classical equations of motion for c-number fields, rather than working directly with the corresponding operator-valued Heisenberg equations. In the Schr\"odinger picture, the state at $N_0$ is
\begin{equation}
    |\Psi(N_0)\rangle
    =
    U(N_0,-\infty)|0_{\rm in}\rangle ,
\end{equation}
with the corresponding wave functional
\begin{equation}
    \Psi_{N_0}[\phi_0]
    \equiv
    \langle \phi_0|\Psi(N_0)\rangle .
\end{equation}
Equivalently, it can be represented by the path integral
\begin{equation}
    \Psi_{N_0}[\phi_0]
    =
    \int_{\substack{
    \text{past boundary}\\
    \phi(N_0)=\phi_0
    }}
    {\cal D}\phi\,
    \exp\left[
        \frac{i}{\hbar}
        S_{-\infty,N_0}[\phi] 
    \right] \equiv \exp\left[
        \frac{i}{\hbar}W_0[\phi_0]
    \right].
\end{equation}
where $W_0[\phi_0]$ contains all information about the evolution from the asymptotic past to $N_0$. The state at $N_0$ therefore need not coincide with the instantaneous vacuum at $N_0$; it is instead the state obtained by evolving the asymptotic in-vacuum up to this time. Then, the propagation of the state from $N_0$ to $N$ is described by
\begin{align}
    \Psi_N[\phi_N]
    ={}&
    \int {\cal D}\phi_0\,
    \Psi_{N_0}[\phi_0]
    \int_{\substack{
    \phi(N_0)=\phi_0\\
    \phi(N)=\phi_N
    }}
    {\cal D}\phi\,
    \exp\left[
        \frac{i}{\hbar}
        S_{N_0,N}[\phi]
    \right],
    \label{eq:wavefunctional-propagation}
\end{align}
where $\phi_N(\mathbf x)$ denotes the field configuration on the final hypersurface. Here $S_{N_0,N}$ denotes the full action, including possible boundary terms, which may be written as
\begin{equation}
    S_{N_0,N}[\phi]
    =
    S_{\mathrm{bulk},N_0,N}[\phi]
    +
    F_{\rm b}(N)
    -
    F_{\rm b}(N_0).
\end{equation}
The equal-time correlators at $N$ are evaluated with the evolved quantum state,
\begin{equation}
    \left\langle
        {\cal O}[\phi(N)]
    \right\rangle
    =
    \frac{
    \displaystyle
    \int{\cal D}\phi_N\,
    \left|\Psi_N[\phi_N]\right|^2
    {\cal O}[\phi_N]
    }{
    \displaystyle
    \int{\cal D}\phi_N\,
    \left|\Psi_N[\phi_N]\right|^2
    },
    \label{eq:tree-correlator-final-state}
\end{equation}

Now let us consider the tree level by taking the saddle point of the path integral and see a computational method of correlators using EoMs. At tree level, Eq.~\eqref{eq:wavefunctional-propagation} is evaluated by the saddle-point method. For a prescribed final configuration $\phi_N$, the saddle is the classical solution $\phi_{\rm cl}$ satisfying $\phi_{\rm cl}(N)=\phi_N$ and the state-dependent initial condition at $N_0$ derived below. Denoting the corresponding initial configuration by $\phi_0^\star$, the final wave functional is therefore
\begin{equation}
    \Psi_N^{\rm tree}[\phi_N]
    \simeq
    {\cal N}_N
    \exp\left[
        \frac{i}{\hbar}
        \left\{
        W_0[\phi_0^\star]
        +
        S_{N_0,N}[\phi_{\rm cl}]
        \right\}
    \right].
    \label{eq:tree-final-wavefunctional}
\end{equation}
The final configuration $\phi_N$ is fixed in constructing $\Psi_N[\phi_N]$, whereas the initial configuration $\phi_0$ is integrated over and is therefore varied at the saddle. Therefore, $\delta\phi_0(\mathbf x)$ is arbitrary for a variational principle while  the boundaries are fixed $\delta\phi(N)=\delta\phi(-\infty)=0$. The saddle-point variation is consequently
\begin{align}
    0&=\delta
    \left[
        S_{N_0,N}[\phi]
        +
        W_0[\phi_0]
    \right] \notag\\
    &= \int \dd N'\dd x^3 \frac{\delta S_{N_0,N}}
         {\delta\phi(N',\mathbf x)}\delta\phi(N',\mathbf{x}) + \int d^3x\,
    \com{    \frac{\delta W_0[\phi_0]}
         {\delta\phi_0(\mathbf x)}-\Pi_{\rm c}(N_0,\mathbf x)}
    \delta\phi_0(\mathbf x),
    \label{eq:full-saddle-variation}
\end{align}
where $\Pi_{\rm c}$ is the momentum canonically conjugate to $\phi$ as defined from the full action, including the contribution of any boundary term. This equation can be interpreted to be the classical equation of motion,
\begin{equation}
    \frac{\delta S_{N_0,N}}
         {\delta\phi(N',\mathbf x)}
    =0,
    \qquad
    N_0<N'<N.
\end{equation}
with the initial hypersurface quantity:
\begin{equation}
    \Pi_{\rm c}(N_0,\mathbf x)
    =
    \frac{\delta W_0[\phi_0]}
         {\delta\phi_0(\mathbf x)}.
    \label{eq:initial-state-boundary-condition}
\end{equation}
Thus, this relation is not obtained by varying a problem in which the initial boundary configuration is fixed. Rather, the quantum state supplies a weight for all possible configurations $\phi_0$ in Eq.~\eqref{eq:wavefunctional-propagation}, and the saddle point of this boundary integration determines the relation between the initial field and its canonical momentum. Together with the prescribed final configuration, this relation determines the classical saddle-point solution. Since $W_0$ is generally complex, the corresponding saddle-point configurations need not be real. In particular, for a Gaussian state, $    W_0[\phi_0]
    =
    \frac12
    \int d^3x\,d^3y\,
    \phi_0(\mathbf x)
    K_0(\mathbf x,\mathbf y)
    \phi_0(\mathbf y)$
where $K_0(\mathbf x,\mathbf y)$ is the generally complex Gaussian kernel characterizing the state at $N_0$. Since $W_0$ contains no cubic term, its functional derivative contains no term quadratic in the linear perturbation, which satisfies $\Pi_{\rm c}(N_0,\mathbf x)
    =
    \int d^3y\,
    K_0(\mathbf x,\mathbf y)
    \phi_0(\mathbf y)$. Thus, the Gaussian initial state does not itself introduce an independent quadratic contribution to the initial canonical momentum. This property will be used below when determining the nonlinear initial conditions in our superhorizon setup.\footnote{Since we focus only on the superhorizon evolution and neglect the nonlinear dynamics prior to the initial hypersurface, the state used in our concrete setup will be approximated as Gaussian at $N_0$.} In summary, the relation between this quantum state and the classical initial data follows directly from the saddle-point representation of the path integral. Once the initial data are fixed in this way, the classical equations of motion determine the subsequent evolution from $N_0$ to $N$. 

\subsection{Application to slow-roll attractor models}
We now demonstrate explicitly how the classical equation of motion, together with the initial conditions determined by the quantum state, can be used to reproduce the tree-level bispectrum in slow-roll attractor models. We take $N_{\rm h}$ as the initial hypersurface of the superhorizon description, corresponding to $N_0$ in the general discussion above, and identify the field variable there with $\delta\phi/\bar\pi$. Since nonlinear interactions prior to $N_{\rm h}$ are neglected by construction, we approximate the state inherited at this hypersurface as Gaussian. The nonlinear evolution for $N>N_{\rm h}$ is then entirely generated by the superhorizon action. The initial state is approximated by the Gaussian wave functional
\begin{equation}
\Psi_{\rm h}
\left[
\left(\frac{\delta\phi}{\bar\pi}\right)_{\rm h}
\right]
=
{\cal N}_{\rm h}
\exp
\left[
\frac{i}{\hbar}
W_{\rm h}
\left[
\left(\frac{\delta\phi}{\bar\pi}\right)_{\rm h}
\right]
\right],
\label{eq:Gaussian-initial-wavefunctional-bispectrum}
\end{equation}
where ${\cal N}_{\rm h}$ is a field-independent normalization and
$W_{\rm h}$ is the quadratic phase functional characterizing the Gaussian state at $N_{\rm h}$, which is the particular realization of $W_0$ introduced above. According to the general argument above, the initial state satisfies Eq.~\eqref{eq:initial-state-boundary-condition}:
\begin{equation}
{
\frac{\delta W_{\rm h}}
{\delta
(\delta\phi/\bar\pi)_{\rm h}(\mathbf{x})} =\Pi^{}_{\rm c}(N_{\rm h},\mathbf{x})
= 2\mpl^2a^3\epsilon_1H
\frac{d}{dN}
\left(
\frac{\delta\phi}{\bar\pi}
\right)
+
\frac{\mpl^2a^3\epsilon_1H}{2}
\epsilon_1
\left(
2\epsilon_1-3\epsilon_2
\right)
\left(
\frac{\delta\phi}{\bar\pi}
\right)^2 
},
\label{eq:saddle-boundary-condition}
\end{equation}
where $\Pi_{\rm c}$ is the momentum canonically conjugate to $\delta\phi/\bar\pi$ defined from the full superhorizon action up to cubic order, including $S_{\rm b}$.
The second term originates from the boundary contribution $S_{\rm b}$ and is therefore automatically included in the canonical initial condition
\eqref{eq:saddle-boundary-condition}. Let us now specify the initial conditions for the second-order perturbations. Since $W_{\rm h}$ is quadratic, its functional derivative is linear in the initial field and contains no term quadratic in the first-order perturbation:
\begin{equation}
\left(
\frac{\delta\phi^{(2)}}{\bar\pi}
\right)
(N_{\rm h},\mathbf{x})
=0.
\label{eq:second-order-initial-field}
\end{equation}
Together with the Gaussian boundary condition, this choice implies that the second-order canonical momentum also vanishes:
\begin{equation}
\Pi_{\rm c}^{(2)}(N_{\rm h},\mathbf{x})=2\mpl^2a^3\epsilon_1H
\dif{}{N}
\left(
\frac{\delta\phi^{(2)}}{\bar\pi}
\right)
+
\frac{\mpl^2a^3\epsilon_1H}{2}
\epsilon_1
\left(
2\epsilon_1-3\epsilon_2
\right)
\left(
\frac{\delta\phi^{(1)}}{\bar\pi}
\right)^2 =0.
\label{eq:second-order-canonical-initial-condition}
\end{equation}
Thus, although the second-order canonical momentum vanishes for the Gaussian initial state in the present approximation, the initial derivative of
$(\delta\phi/\bar\pi)^{(2)}$ does not vanish because the canonical momentum contains the nonlinear contribution generated by the boundary term. 

Equations~\eqref{eq:second-order-initial-field} and
\eqref{eq:second-order-canonical-initial-condition} provide the initial data for the second-order classical solution. The second-order equation of motion obtained from the corresponding quadratic and cubic superhorizon actions is
\begin{align}
&
\dif{}{N}
\left[
\mpl^2a^3\epsilon_1H
\dif{}{N}
\left(
\frac{\delta\phi}{\bar\pi}^{(2)}
\right)
\right]
=
\frac14
\dif{}{N}
\left(
\mpl^2a^3\epsilon_1H
\epsilon_2\epsilon_3
\right)
\left(
\frac{\delta\phi^{(1)}}{\bar\pi}
\right)^2 .
\label{eq:second-order-EOM-bispectrum}
\end{align}
Since the growing linear superhorizon solution satisfies
\begin{equation}
\frac{d}{dN}
\left(
\frac{\delta\phi^{(1)}}{\bar\pi}
\right)
=0,
\end{equation}
the source in Eq.~\eqref{eq:second-order-EOM-bispectrum} can be integrated directly. Note that the form of Eq.~\eqref{eq:second-order-EOM-bispectrum} is exactly the same as that of Eq.~\eqref{eq:second-order-attractor}. Using
Eqs.~\eqref{eq:second-order-initial-field} and \eqref{eq:second-order-canonical-initial-condition}, one obtains
\begin{align}
\left(
\frac{\delta\phi^{(2)}}{\bar\pi}
\right)
={}&
\frac{
\epsilon_2(N)-\epsilon_2(N_{\rm h})
}{4}
\left(
\frac{\delta\phi^{(1)}}{\bar\pi}
\right)^2
-
\frac{
\mpl^2a^3(N_{\rm h})
\epsilon_1(N_{\rm h})
H(N_{\rm h})
}{4}
\left\{
\epsilon_2(N_{\rm h})
\epsilon_3(N_{\rm h})
\right.
\nonumber\\
&\left.
\qquad\qquad
+
\epsilon_1(N_{\rm h})
\left[
2\epsilon_1(N_{\rm h})
-
3\epsilon_2(N_{\rm h})
\right]
\right\}
\left[
\int_{N_{\rm h}}^{N}
\frac{dN_2}
{\mpl^2a^3\epsilon_1H}
\right]
\left(
\frac{\delta\phi^{(1)}}{\bar\pi}
\right)^2 .
\label{eq:second-order-solved-field}
\end{align}

We now relate this classical solution~\eqref{eq:second-order-solved-field} to the tree-level quantum correlator. As established in the general formulation, the tree-level wave functional on the final hypersurface is obtained by evaluating the action and the initial-state functional on the corresponding classical saddle-point solution. The equal-time three-point function is then computed from the probability functional of the evolved state:
\begin{align}
&
\left\langle
\delta\phi(N,\mathbf{x}_1)
\delta\phi(N,\mathbf{x}_2)
\delta\phi(N,\mathbf{x}_3)
\right\rangle_{\rm tree}
=
\frac{
\displaystyle
\int
{\cal D}
\left(\frac{\delta\phi}{\bar\pi}\right)_N
\left|
\Psi_N^{\rm tree}
\left[
\left(\frac{\delta\phi}{\bar\pi}\right)_N
\right]
\right|^2
\prod_{a=1}^{3}
\delta\phi(N,\mathbf{x}_a)
}{
\displaystyle
\int
{\cal D}
\left(\frac{\delta\phi}{\bar\pi}\right)_N
\left|
\Psi_N^{\rm tree}
\left[
\left(\frac{\delta\phi}{\bar\pi}\right)_N
\right]
\right|^2
}.
\label{eq:three-point-final-wavefunctional}
\end{align}
The perturbative dependence of the evolved wave functional on the final field configuration $\Psi_N^{\rm tree}$ is determined by the classical saddle-point solution, together with the Gaussian initial-state functional and the full action. Within the superhorizon approximation considered here, the leading connected three-point function can equivalently be obtained from the second-order classical solution~\eqref{eq:second-order-solved-field} by performing Gaussian contractions of the linear perturbations. In particular, the quadratic dependence on \(\delta\phi^{(1)}\) generates the three-point function through the two possible contractions with the remaining linear fields. In this way, the classical equation of motion together with the initial condition fixed by the quantum state reproduces the bispectrum obtained from the direct in-in computation \eqref{B deltaphi IR}.

\section{The late-time IR behavior of the nonlinear extension}\label{App:C}
Let us examine the nonlinear extension of the stochastic formalism demonstrated in Sec.~\ref{sec:nonlinear}, from the perspective of resumming late-time IR divergences which is another motivation for using this framework. As noted above, most studies of stochastic inflation (including the original formulation \cite{Starobinsky:1986fx,Starobinsky:1994bd}) focus exclusively on the IR–IR self-interaction of the field. In our formulation, this corresponds to the contribution labeled ``$\mathrm{a}$", which is expected to dominate in the late-time limit. To verify whether this statement is accurate, we consider the simplified setup of a massless scalar field with a cubic self-interaction, $\left(V(\phi)=\lambda \phi^3\right)$, and compare the three-point function computed using only the contribution labeled ``$\mathrm{a}$'', i.e.\ Eq.~\eqref{second_IR1_solution_3-point_final}, with the total three-point function, which also includes the IR--UV and UV--UV interactions, i.e.\ Eq.~\eqref{second_IR_total_solution_3-point_final}. 

\begin{figure}[h]
    \centering
    \includegraphics[scale=0.3]{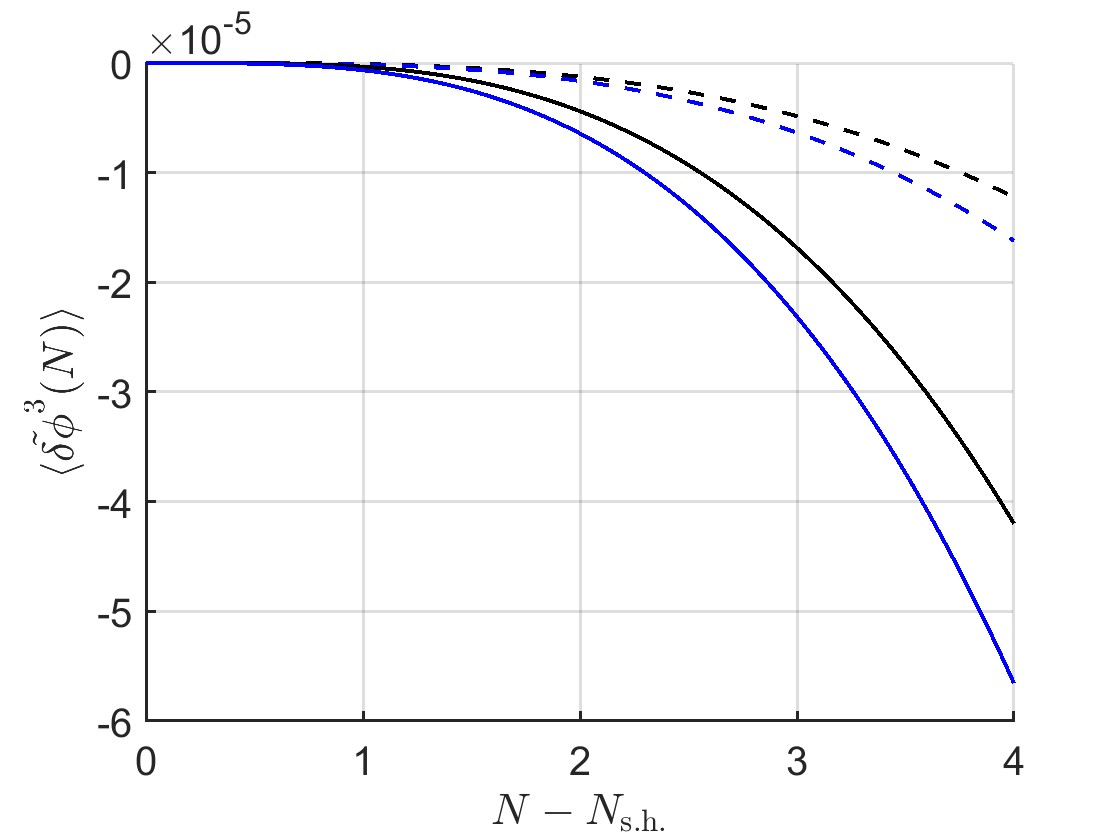}
    \caption{Numerical results for the three-point functions in Eqs.~\eqref{second_IR1_solution_3-point_final} (black dashed line) and \eqref{second_IR_total_solution_3-point_final} (black solid line) at all orders in \(\lambda\). The factor-of-3 difference between the two remains visible. For better visualization, we also include the result at leading order in \(\lambda\) of Eqs.~\eqref{3-point_a_cubic} (blue dashed line) and \eqref{3-point_cubic} (blue solid line). All three point functions are divided by $\mpl^3$ to make them dimensionless. }
    \label{Fig}
\end{figure}

In order to gain analytical understanding we work at leading order in $\lambda$, the background equation of motion in this case is

\begin{equation}
    \dif{\bar\phi}{N}+\lambda\frac{\bar{\phi}^2}{6\bar{H}^2} = 0,
\end{equation}
and the velocity $\bar{\pi}\simeq -\lambda\frac{\bar{\phi}_0}{6\bar{H}^2}$ is a constant. Furthermore, $P_k(N,N_1)$ is given by 
\begin{equation}
    P_k(N,N_1) = \frac{{2\pi^2}}{k^3}\left(\frac{\bar{H}}{2\pi}\right)^2.
\end{equation}

The three-point functions \eqref{second_IR1_solution_3-point_final},\eqref{second_IR_total_solution_3-point_final} at $\mathcal{O}\left(\lambda\right)$ are given by 
\begin{align}
    &\langle(\phi_\uir(N,\mathbf x)
    -\langle\phi_\uir(N,\mathbf x)\rangle)^3\rangle\big|_\mathrm a\simeq-3\frac{\lambda}{\bar{H}^2}\left(\frac{\bar{H}}{2\pi}\right)^4 \int_{N_\mathrm{s.h.}}^{N}\frac{\dd N_1}{a(N_1)^3}\int_{N_\mathrm{s.h.}}^{{N_1}}\dd N_2 a(N_2)^3\left(\int_{\sigma a\left(N_\mathrm{s.h.}\right)\bar{H}}^{\sigma a\left(N_2\right)\bar{H}}\frac{\dd k}{k}\right)^2 \notag\\
    &\quad = \frac{1}{3}\frac{\lambda}{\bar{H}^2}\left(\frac{\bar{H}}{2\pi}\right)^4\left(\frac{2}{9}\left(1-e^{-3\left(N-N_\mathrm{s.h.}\right)}\right)-\frac{2}{3}\left(N-N_\mathrm{s.h.}\right)+\left(N-N_\mathrm{s.h.}\right)^2-\left(N-N_\mathrm{s.h.}\right)^3\right) \notag\\
    &\quad \simeq -\frac{1}{3}\frac{\lambda}{\bar{H}^2}\left(\frac{\bar{H}}{2\pi}\right)^4\left(N-N_\mathrm{s.h.}\right)^3,
    \label{3-point_a_cubic}
\end{align}
and
\begin{align}
    &\langle(\phi_\uir(N,\mathbf x)
    -\langle\phi_\uir(N,\mathbf x)\rangle)^3\rangle\simeq-3\frac{\lambda}{\bar{H}^2}\left(\frac{\bar{H}}{2\pi}\right)^4 \int_{N_\mathrm{s.h.}}^{N}\frac{\dd N_1}{a(N_1)^3}\int_{N_\mathrm{s.h.}}^{N_1}\dd N_2 a(N_2)^3\left(\int_{\sigma a\left(N_\mathrm{s.h.}\right)\bar{H}}^{\sigma a\left(N\right)\bar{H}}\frac{\dd k}{k}\right)^2 \notag\\
    &\quad = \frac{1}{3}\frac{\lambda}{\bar{H}^2}\left(\frac{\bar{H}}{2\pi}\right)^4\left(N-N_\mathrm{s.h.}\right)^2\left(\left(1-e^{-3\left(N-N_\mathrm{s.h.}\right)}\right)-3\left(N-N_\mathrm{s.h.}\right)\right) \notag\\
    &\quad \simeq -\frac{\lambda}{\bar{H}^2}\left(\frac{\bar{H}}{2\pi}\right)^4\left(N-N_\mathrm{s.h.}\right)^3,
    \label{3-point_cubic}
\end{align}
where in the last line we are using the large-$N$ limit in order to study late-time IR divergences. In this limit, we can set $ W\prn{\frac{|\mathbf k+\mathbf q|}{k_\sigma(N)}}=1$. The large-\(N\) limit of Eq.~\eqref{3-point_a_cubic} and that of Eq.~\eqref{3-point_cubic} coincide with Eqs.~(18) and (37) of Ref.~\cite{Palma:2023uwo}, respectively. This is not surprising: Eq.~(18) of Ref.~\cite{Palma:2023uwo} relies on the typical assumption \(V\bigl(W\{\phi\}\bigr)\), which accounts only for IR--IR interactions, whereas Eq.~(37) also includes IR--UV and UV--UV interactions by applying the IR window function outside the full potential, i.e.\ \(W\bigl\{V(\phi)\bigr\}\).

Note that although the large-\(N\) behavior of Eqs.~\eqref{3-point_a_cubic},\eqref{3-point_cubic} is the same, their prefactors differ by a factor of 3\footnote{Strictly speaking, a quantitative comparison should also account for subhorizon contributions, in addition to the nonlinear superhorizon evolution considered here.}. This may affect possible ways of regularizing IR divergences, such as the inclusion of counterterms. This suggests that IR--UV and UV--UV interactions can remain relevant even when late-time IR divergences are resummed. Finally, for completeness, we compute numerically the three-point functions in Eqs.~\eqref{second_IR1_solution_3-point_final}, \eqref{second_IR_total_solution_3-point_final} to all orders in \(\lambda\) and show the results in Fig.~\ref{Fig}, where the factor-of-3 difference is seen to persist.

\bibliographystyle{JHEP}
\bibliography{ref}
\end{document}